\newcommand*{\ATLASLATEXPATH}{./}
\pdfoutput=1
\documentclass[cernpreprint, texlive=2016, USenglish]{\ATLASLATEXPATH atlasdoc}
\usepackage{\ATLASLATEXPATH atlaspackage}
\usepackage{\ATLASLATEXPATH atlasbiblatex}

\usepackage{\ATLASLATEXPATH atlasphysics}

\usepackage{units}
\usepackage{multirow}

\usepackage{Multicharge2015Paper-defs}

\AtlasTitle{Search for heavy long-lived multi-charged particles in proton--proton collisions at $\rts=\unit[13]{\TeV{}}$ using the ATLAS detector}
\AtlasAbstract{%
  A search for heavy long-lived multi-charged particles is performed using the ATLAS detector at the LHC. Data with an integrated luminosity of \intlumi{} collected in 2015 and 2016 from proton--proton collisions at $\rts=\unit[13]{\TeV{}}$ are examined. Particles producing anomalously high ionization, consistent with long-lived massive particles with electric charges from ${|q|=2e}$ to ${|q|=7e}$, are searched for. No events are observed, and $95\%$ confidence level cross-section upper limits are interpreted as lower mass limits for a Drell--Yan production model. Multi-charged particles with masses between $\unit[50]{\GeV{}}$ and $980$--$\unit[1220]{\GeV{}}$ (depending on their electric charge) are excluded.
}
\author{The ATLAS Collaboration}
\PreprintIdNumber{CERN-EP-2018-284}
\AtlasJournal{\PRD}

\AtlasJournalRef{\PRD 99 (2019) 052003}
\AtlasDOI{10.1103/PhysRevD.99.052003}

\hypersetup{pdftitle={ATLAS document},pdfauthor={The ATLAS Collaboration}}

\begin{document}

\maketitle

\section{Introduction}
\label{sec:intro}

This article describes a search for heavy long-lived 
multi-charged particles (MCPs) in $\rts=\unit[13]{\TeV{}}$ proton--proton collision data
collected in 2015 and 2016 by the ATLAS detector at the CERN Large Hadron Collider (LHC)~\cite{Evans:2008zzb}. 
The search, conducted on a sample of data corresponding to an integrated luminosity of \intlumi{}, 
is performed in the MCP mass range from $50$ to \unit[1400]{\GeV{}}, for
electric charges\footnote{Wherever a charge is quoted for exotic particles, the charge 
conjugate state is also implied.} 
${|q|=ze}$, with charge numbers $2 \leq z \leq 7$. 
An observation of such particles, possessing an electric charge above the elementary
charge $e$, would be a signature for physics beyond the Standard Model (SM).
Several theoretical models predict such particles. 
AC-leptons, as predicted by the almost-commutative model~\cite{Stephan:2005uj}, 
are pairs of SU(2) electroweak singlets with opposite electromagnetic charges and no 
other gauge charges of the SM, which makes them behave as heavy stable charged leptons. 
Technibaryons, predicted by the walking-technicolor model~\cite{Sannino:2004qp}, are 
Goldstone bosons made of two techniquarks or two anti-techniquarks with 
an arbitrary value of the electric charge. The lightest technibaryon is expected to be stable in 
the absence of processes violating the technibaryon number conservation law. 
Doubly charged Higgs bosons are predicted by the left--right symmetric model~\cite{Mohapatra:1974hk}
in Higgs triplets in a model postulating a right-handed 
version of the weak interaction. Its gauge symmetry is spontaneously broken at a high mass scale, 
leading to parity-violation in the weak-interaction sector of the SM. 
Only leptonic decay modes would be characteristic of such particles, as shown in the model 
described in \Ref{\cite{Aulakh:1997ba}}. The $H^{\pm\pm} \rightarrow W^{\pm}W^{\pm}$ decays are assumed to be 
suppressed. 
The supersymmetric left-right model~\cite{Aulakh:1997ba}, which imposes lepton number conservation, predicts a 
light $H^{\pm\pm}$ boson with null lepton number, forbidding its decays to two same-sign leptons and making the 
$H^{\pm\pm}$ boson long-lived.
Any observation of the particles predicted by the first two models could have
implications for the formation of composite dark matter: 
the doubly charged particles (or, in general, particles with an even charge ${|q| = 2ne}$)
could explain some excesses (e.g., positron excess) observed in direct and indirect searches 
for dark matter~\cite{Khlopov:2011tn,Belotsky:2015gsa}.
Particles with half-integer charge are considered in this search in order to allow  
continuous mass limits to be set between the $2e$ and $7e$ cases.
So far, no such particles have been observed in cosmic-ray~\cite{Cecchini:2008su} or collider searches, 
including several recent searches at the Tevatron~\cite{Acosta:2005np} and 
the LHC~\cite{Aad:2015oga,Chatrchyan:2013oca,Khachatryan:2016sfv}.

A purely electromagnetic coupling, proportional to the electric charge of the MCPs, is assumed for the production model. 
In this search the MCPs are assumed to live long enough to traverse the entire ATLAS detector without decaying, 
and thus the analysis exploits their muon-like signature, making the muon trigger a natural choice. 
They are highly ionizing, and thus generate an abnormally large ionization signal, \dEdx{}, which leads to 
their significant slowdown. 
Especially for MCPs with the highest charge and lowest mass values, this causes an event to not be triggered and/or 
MCPs to fail reconstruction as muons (this is the main reason for the search to be limited by the $z=7$ MCPs from above). 
The addition of the missing-transverse-momentum trigger mitigates the first issue because a difference in energy 
deposited by the two MCPs in the calorimeter will lead to a non-zero \ET{} vector sum and will make this trigger fire. 
Also, this trigger accepts events with high-mass MCPs that are too slow to fall within the muon-trigger timing window. 
The offline analysis searches for muon-like tracks with high \dEdx{} values in several subdetector systems. 
The background expected from the SM processes (largely high-\pT{} muons) is estimated using a data-driven technique.

\section{ATLAS detector}
\label{sec:detector}

The ATLAS detector~\cite{PERF-2007-01} covers nearly the entire solid angle around the collision point.\footnote{ATLAS 
uses a right-handed
coordinate system with its origin at the nominal interaction point (IP) in the center of the detector and
the $z$-axis along the beam pipe. The $x$-axis points from the IP to the center of the LHC ring, and
the $y$-axis points upward. Cylindrical coordinates ${(r,\phi)}$ are used in the transverse plane,
$\phi$ being the azimuthal angle around the $z$-axis. The pseudorapidity is defined in terms of the polar
angle $\theta$ as ${\eta=-\ln\tan(\theta/2)}$. Angular distance is measured in units
of $\Delta R \equiv \sqrt{(\Delta\eta)^{2} + (\Delta\phi)^{2}}$.} 
The inner tracking detector (ID) consists of a silicon pixel detector (pixel), a silicon
microstrip detector (SCT) and a transition radiation tracker (TRT).
The pixel detector was upgraded in 2014 with the insertion of an additional layer,
the insertable B-layer (IBL)~\cite{Abbott:2018ikt}, mounted on a new beam pipe of smaller diameter.
The pixel detector provides at least four precise space-point measurements per track.
At normal incidence, the average charge released by a minimum-ionizing particle (MIP) in a pixel sensor is 
$\approx 20 000~e^{-}$ ($\approx 16 000~e^{-}$ for the IBL) and the charge threshold is set to 
$3500~e^{-}$ ($2500~e^{-}$ for the IBL)~\cite{Aaboud:2016dgf}. Signals are accepted if they are larger than this threshold. 
The time interval with the signal above the threshold is approximately proportional to the ionization charge
and its dynamic range corresponds to $8.5$ times ($1.5$ times for the IBL) the average charge released by a MIP 
if its track is normal to the silicon detectors and it deposits all its ionization charge in a single pixel. 
If this value is exceeded in the IBL, the electronics signals an excess with an overflow bit; if it is exceeded in 
the other three layers of the pixel detector, the hit information is not recorded due to electronics limitations 
(nor is the fact of the overflow). However, since the charge released by a particle crossing the pixel detector is 
rarely contained within just one pixel, the neighboring pixels preserve the spatial information of this hit.
The SCT consists of four double-layer silicon sensors with binary readout architecture, 
each with a small stereo angle, typically providing eight measurements per track.
The TRT, covering the pseudorapidity range ${\abseta<2.0}$, is a straw-tube
tracking detector capable of particle identification via transition-radiation and 
ionization-energy-loss measurements~\cite{ATLAS:2011qha}. A particle typically crosses $32$~straws.
Discriminators are used to compare the signal from a straw with a low threshold and a high
threshold (HT). 
The HT is designed to discriminate between energy depositions from transition-radiation 
photons and the energy loss of MIPs.
Roughly three times the energy deposition of a MIP is needed to generate an HT hit.
MCPs would produce a large number of HT hits along their trajectories due to their high ionizing power.

The ID is surrounded by a thin superconducting solenoid providing a \unit[2]{T} axial magnetic field,
and by a high-granularity lead/liquid-argon (LAr) sampling electromagnetic calorimeter.
An iron/scintillator tile calorimeter provides hadronic-energy measurements in the central pseudorapidity region.
The endcap and forward regions are instrumented with LAr electromagnetic and hadronic calorimeters.
The calorimeter system is surrounded by a muon spectrometer (MS) incorporating three superconducting toroidal
magnet assemblies. The MS is instrumented with tracking detectors designed to measure the momenta of muons. 
Resistive-plate chambers (RPC) in the barrel
region (${\abseta < 1.05}$) and thin-gap chambers (TGC) in the endcap
regions (${1.05 < \abseta < 2.4}$) provide signals for the trigger.
Monitored-drift-tube (MDT) chambers typically provide $20$--$25$ hits per crossing track in the pseudorapidity
range ${{\abseta < 2.7}}$, from which a high-precision momentum measurement is derived.
In each amplifier--shaper--discriminator channel, an analog-to-digital converter is used to measure
the signal charge in the \unit[18.5]{ns} integration gate following the initial threshold crossing~\cite{Arai:2008zzb}.
Cathode-strip chambers complement the tracking capabilities of the MDTs in the high-rate forward regions. 

A two-level trigger system is used to select interesting events~\cite{Aaboud:2016leb}.
The first trigger level is implemented in hardware and uses a subset of the detector information
to reduce the event rate to a design value of at most \unit[100]{kHz}.
This is followed by the software-based high-level trigger, which reduces the event rate to about \unit[1]{kHz}.
 
The amount of material in the ID varies from one-half to two radiation lengths.
The overall amount of material traversed by an MCP up to the last measurement surface, which includes the calorimeters and the MS,
may be as high as $75$ radiation lengths. Muons typically lose \unit[3]{\GeV{}} penetrating the calorimeter system.
The energy loss for MCPs with charge $z$ would be $z^{2}$ times this value, i.e. up to \unit[150]{\GeV{}} for ${z=7}$.
 
The muon transverse momentum measured by the MS after the energy loss in the calorimeters is denoted by $\pT^{\mu}$, 
while transverse momentum of charged particles measured by the ID or the combination of the ID and MS is denoted by $\pT{}$.
Charged-particle trajectories are reconstructed using standard algorithms. 
Since these algorithms assume particles with unit electric charge, 
the momenta of MCPs are underestimated by a factor $z$, as the track curvature is proportional to ${\pT / z}$.

\section{Samples of simulated events}
\label{sec:McSamples}

Benchmark samples of simulated events with MCPs were generated
for a mass of \unit[50]{\GeV{}} and for a range of masses between $200$ and \unit[1400]{\GeV{}}
in steps of \unit[200]{\GeV{}}, for charges $ze$ with ${z=2}$, $2.5$, \ldots, $7$.
Lepton-like pairs of MCPs were generated via the lowest-order Drell--Yan (DY) process implemented in
\MGMCatNLO{}~2.3.3~\cite{Alwall:2014hca} with only photon exchange included. 
This implementation of the DY production process models the kinematic 
distributions and determines the cross-sections. 
Cross-section values for MCP pair production range from hundreds of picobarns (mass of \unit[50]{\GeV{}}, ${z=7}$) 
down to a hundredth of a femtobarn (mass of \unit[1400]{\GeV{}}, ${z=2}$).
Events were generated using the NNPDF23LO~\cite{Carrazza:2013axa} parton
distribution functions with the A14 set of tuned parameters~\cite{ATL-PHYS-PUB-2014-021}, and 
\PYTHIAV{8.205}~\cite{Sjostrand:2014zea, Sjostrand:2006za} was
used for hadronization and underlying-event generation.

Simulated samples with muons from \Zmm{} decays were generated using \POWHEGBOX{} v2~\cite{Nason:2004rx,Alioli:2008gx}
interfaced to the \PYTHIAV{8.186} parton shower model. The AZNLO tuned parameters~\cite{AZNLO:2014} 
were employed, with the CTEQ6L1 PDF set~\cite{Pumplin:2002vw} for the modeling of non-perturbative effects.
The {\normalfont \scshape EvtGen}~1.2.0 program~\cite{EvtGen} was utilized for the properties of $b$- and $c$-hadron decays. 

A full \GEANT{}4 simulation~\cite{Agostinelli:2002hh, Aad:2010ah} was used to model the response of the ATLAS detector.
Each simulated hard-scattering event was overlaid with simulated minimum-bias
events (``pileup'') generated with \PYTHIA in order to reproduce the observed
distribution of the number of proton--proton collisions per bunch crossing.
The simulated events are reconstructed and analyzed in the same way as the experimental data.

\section{Event and candidate selections}
\label{sec:Selection}

The search relies on the ionization energy released by high-charge particles and measured 
in the pixel, TRT, and MDT subdetector systems. Acceptance is restricted to 
the pseudorapidity range ${\abseta < 2.0}$ because of the TRT geometrical limitation.

The selection is logically divided into four steps: trigger and event selection,
preselection, tight selection, and final selection.
While the first two steps rely on muon and missing-transverse-momentum (\met{}) signals as well as event topology, the tight and 
final selection steps rely on the ionization estimators not available at the trigger level.
These estimators are introduced later in this section.
An event is considered to be a candidate event if it has at least one candidate 
MCP (a reconstructed particle, which satisfies all selection criteria).

\subsection{Trigger and event selections}
\label{sec:TrigAndEvSel}

Events collected in 2015 and 2016 with a single-muon trigger with no isolation requirement and 
a transverse-momentum threshold
of ${\pT/z = \unit[50]{\GeV{}}}$ are considered.
This trigger is only sensitive to particles with velocity $\beta=v/c>0.6$ due to a timing window, 
within which particles must reach the MS, which limits the trigger efficiency.

To compensate for inefficiencies in the single-muon trigger, an additional calorimeter-based trigger that imposes 
a threshold on the magnitude of the \etmiss{} was employed. 
The \etmiss{} threshold was $\unit[70]{\GeV{}}$ in 2015 and was raised twice in 2016, first to $\unit[90]{\GeV{}}$ and later to $\unit[110]{\GeV{}}$.
Particles reconstructed in the MS are not accounted for in the trigger \etmiss{} calculation, 
which only takes into account energy deposited in the calorimeters. 
Large missing transverse momentum can be due to a major difference between the energy deposited by the 
two MCPs in the calorimeter leading to a non-zero \ET{} vector sum (because of significant random fluctuations of 
the deposited energy) and also due to an MCP--MCP system recoiling against a jet, 
given that the energy deposited by the MCPs in the calorimeter would not balance the jet energy.

If an event is selected by both of these triggers,
it is assigned to the single-muon trigger for the following analysis.
On average, the exclusive contribution of the \etmiss{} trigger is about $20\%$ of  
the overall number of triggered signal events.

\subsection{Candidate track preselection}
\label{sec:CandTrackPreselAndIonisEstim}

Each candidate track is required to be a ``combined'' muon, i.e.\ reconstructed by combining track segments in 
the ID with those in the MS. These candidate muons must satisfy the ``medium'' criteria defined 
in \Ref{\cite{Aad:2016jkr}}, have ${\pT^{\mu}/z > \unit[50]{\GeV{}}}$, and fall within 
the acceptance region of the TRT (${\abseta < 2.0}$). 

In order to reduce the background of high-ionization signals from two or more
tracks firing the same TRT straws or MDT tubes, each candidate is required
not to have any adjacent tracks with ${\pT/z > \unit[0.5]{\GeV{}}}$ within ${\dR < 0.01}$.

\subsection{Ionization estimators and tight/final selections}
\label{sec:TightAndFinalSels}

The definitions of the tight and final selections require the introduction of ionization estimators.

The average specific energy loss, \dEdx{}, is described by the Bethe--Bloch
formula~\cite{1930AnP...397..325B}. Since a particle's energy loss increases quadratically with its charge,
an MCP would leave a very characteristic signature of high ionization in the detector. Estimates of
\dEdx{} are evaluated for the pixel, TRT and MDT subdetector systems.
The pixel \dEdx{} is calculated from the truncated mean of the \dEdx{} values of the clusters associated with the track, 
excluding the largest (one or two) \dEdx{} measurements. 
The TRT \dEdx{} is the truncated mean of the hit-level \dEdx{} estimates, derived from the time interval 
when the signal remains above the low threshold. 
Each drift tube of the MDT system provides a signal proportional to the charge from ionization; 
a truncated mean of these measurements is treated as the MDT \dEdx{} estimator. 
Apart from the mentioned tail truncations, calibrations and corrections of these \dEdx{} estimators include 
removal of their dependencies on geometrical quantities (pseudorapidity, distance between a particle track 
and an anode wire for the TRT and MDT) and of those related to detector effects: dependence on the number of 
hits, radiation damage leading to run-by-run response difference for the pixel detector, detector occupancy 
for the pixel detector and TRT, difference between the response in the different detector sections for the MDT, etc.

The significance of the \dEdx{} variable in each subdetector is defined by comparing the observed signal,
$\text{\dEdx{}}$, with the average value for a highly relativistic muon:
 
\begin{equation*}
	S(\dEdx{}) = \frac{\text{\dEdx{}} - \langle\text{\dEdx{}}\rangle_{\mu}}{\sigma(\text{\dEdx{}})_{\mu}}.
\end{equation*}

Here $\langle\text{\dEdx{}}\rangle_{\mu}$ and $\sigma(\text{\dEdx{}})_{\mu}$
represent, respectively, the mean and the root-mean-square width of the \dEdx{} distribution for such muons in data.
To calculate these two parameters, a control sample of muons was obtained from $\Zmm{}$ events. 
The muon selection is the same as in the analysis selection discussed in \Sect{\ref{sec:CandTrackPreselAndIonisEstim}}. 
Also, muons are required to belong to an oppositely charged pair with dimuon mass
between \unit[81]{\GeV{}} and \unit[101]{\GeV{}}. These requirements effectively
suppress muons from other processes. 

In addition to the \dEdx{} estimates, the number of IBL clusters with at least one hit in overflow 
(called in the rest of the paper, for simplicity, the number of overflowing IBL clusters) and 
the fraction of HT TRT hits ($f^{\textrm{HT}}$) 
are estimators of the energy loss and are used in the tight selection.

\begin{figure}[ht]
 \begin{center}
   \includegraphics[width=0.6\textwidth]{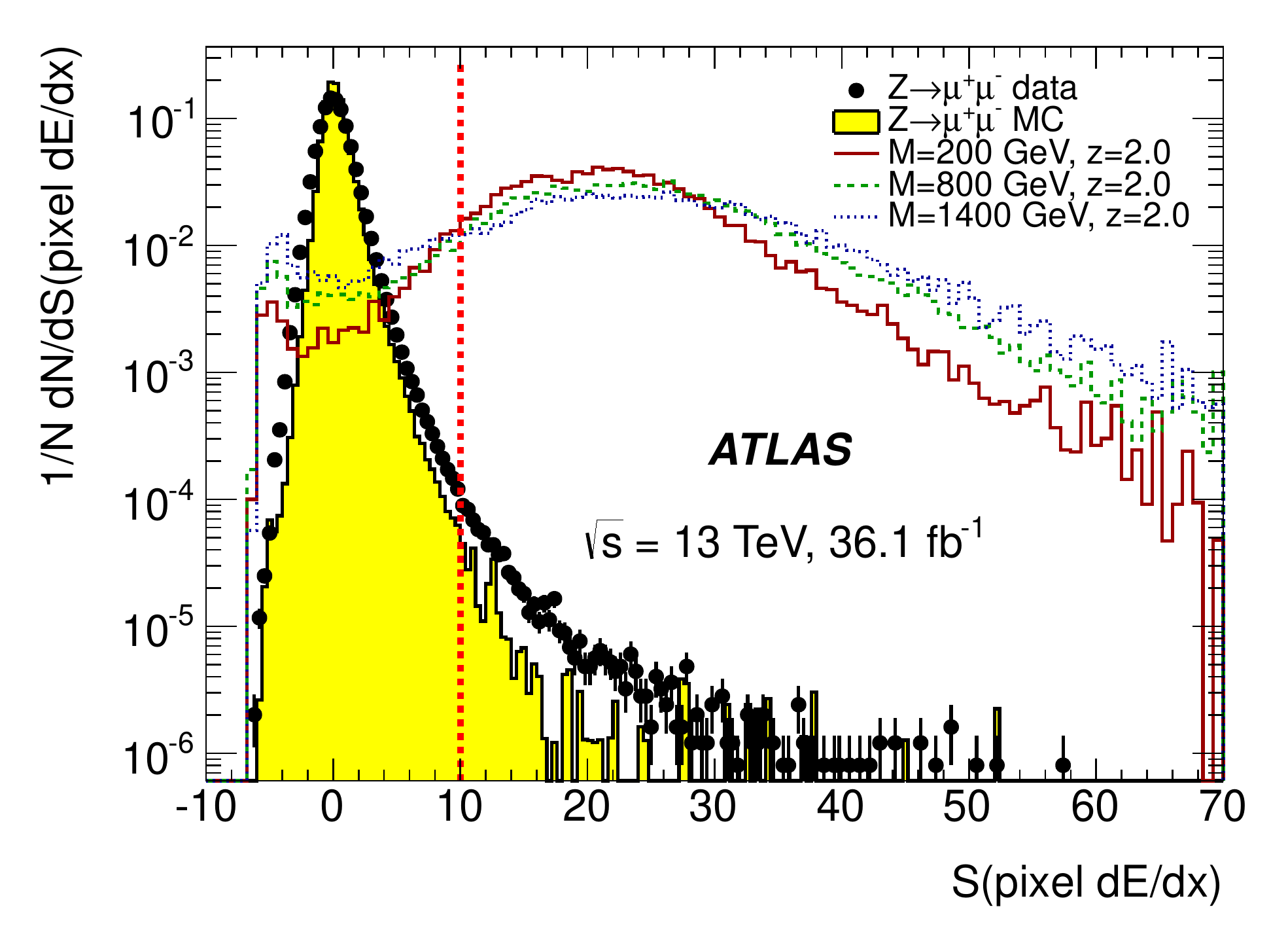}
 \end{center}
 \caption{Normalized distributions of the \dEdx{} significance in the pixel system,
   $S$(pixel \dEdx{}) for muons from $\Zmm{}$ events (data and simulation) and for simulated MCPs 
   passing the preselection requirements.
   Signal distributions are shown for ${z=2}$ and masses of $200$, $800$, and \unit[1400]{\GeV{}}.
   The red dotted line indicates the threshold of the selection criterion for the ${z=2}$ search case.}
 \label{fig:SPixeldEdx}
\end{figure}

\begin{figure}[ht]
 \begin{center}
   \includegraphics[width=0.49\textwidth]{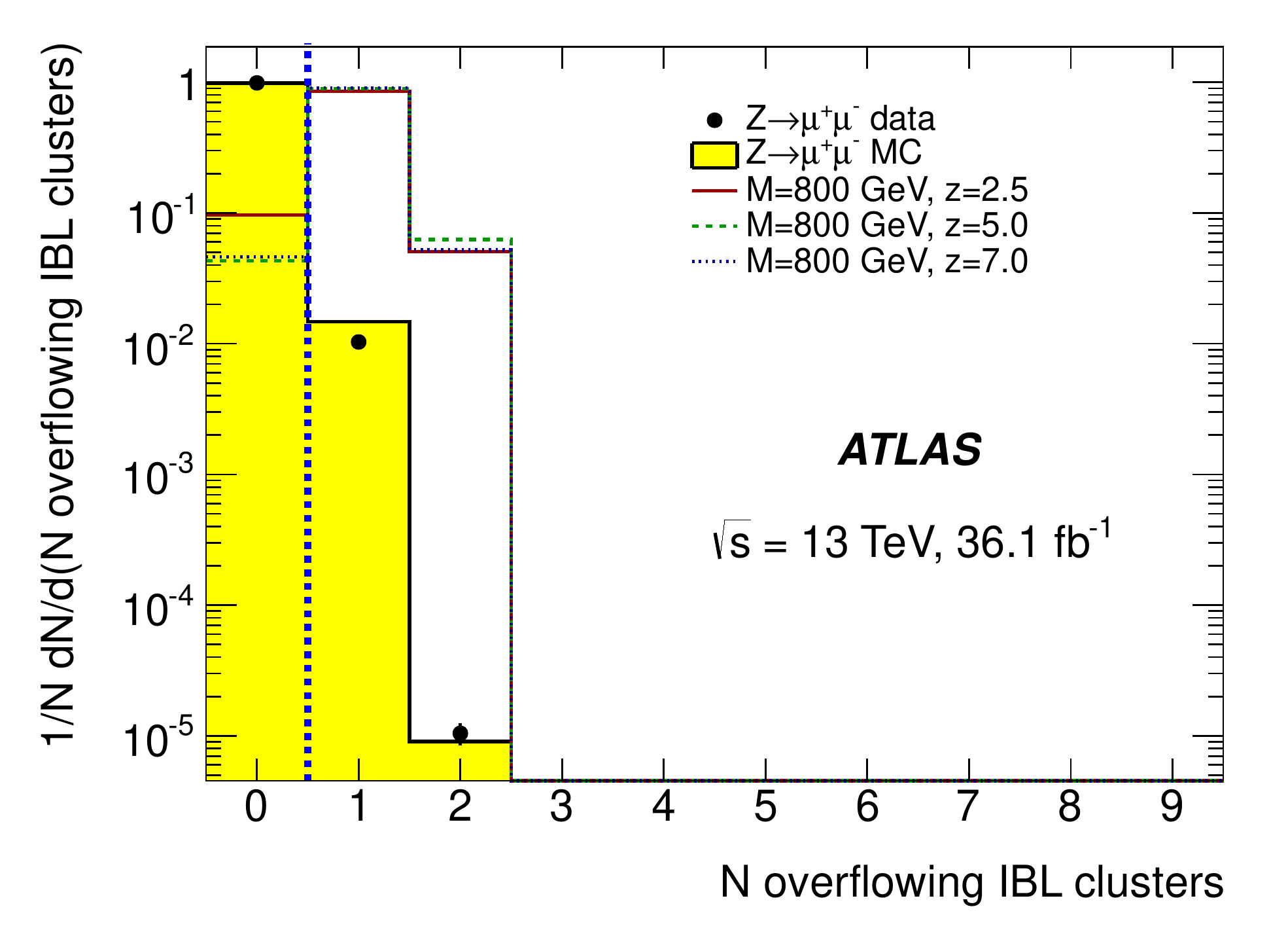}
   \includegraphics[width=0.49\textwidth]{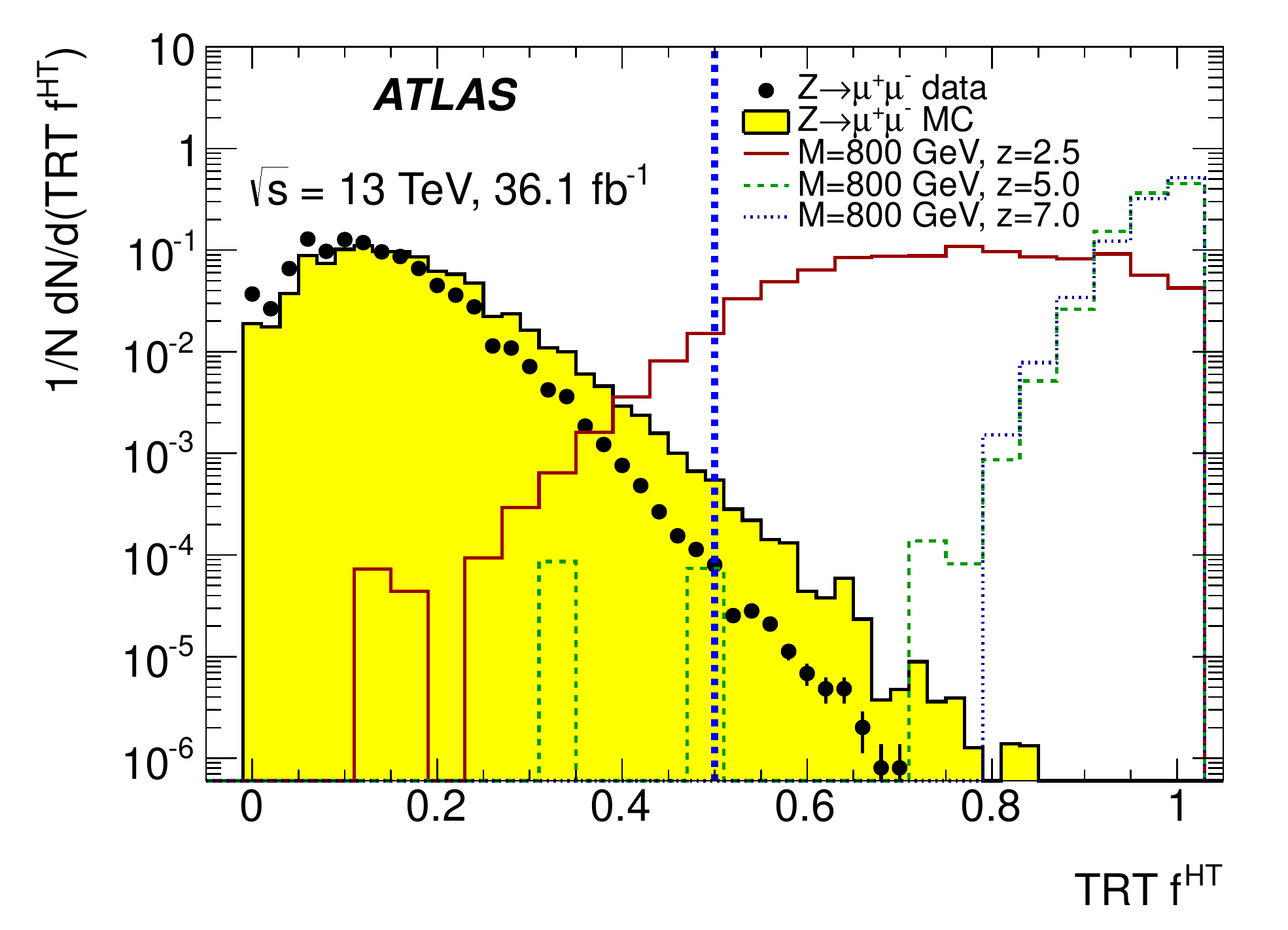}
 \end{center}
 \caption{Normalized distributions of the number of overflowing IBL clusters (left) and
   $f^{\textrm{HT}}$ (right) for muons from $\Zmm{}$ events (data and simulation) and for simulated MCPs 
   passing the preselection requirements.
   Signal distributions are shown for ${z=2.5}$, $5.0$, $7.0$ and a mass of \unit[800]{\GeV{}}.
   The blue dotted lines indicate the thresholds of the selection criteria for the ${z>2}$ search case.}
 \label{fig:NIblOverflownClustersAndTrtFht}
\end{figure}

As seen in \Fig{\ref{fig:SPixeldEdx}}, $S$(pixel \dEdx{}) is a powerful discriminator for 
particles with ${z=2}$. 
The signal region of the tight selection is defined by requiring $S$(pixel \dEdx{}) greater than $10$.
For higher values of $z$, the pixel readout saturates and the corresponding hits are not recorded. 
Therefore, to search for particles with ${z>2}$, the number of overflowing IBL clusters and $f^{\textrm{HT}}$
(see \Fig{\ref{fig:NIblOverflownClustersAndTrtFht}}) are used as discriminating
variables instead, with the signal regions of the tight selection defined by requiring at least one overflowing IBL cluster
and $f^{\textrm{HT}}$ to be above $0.5$. 

For both the ${z=2}$ and ${z>2}$ search cases, the tight selection criteria reduce the background 
contribution (mainly from the high-\pT muons) by at least three orders of magnitude,
while keeping the signal efficiency above $90\%$ relative to the efficiency obtained in the previous selection step.

\begin{figure}[ht]
 \begin{center}
   \includegraphics[width=0.49\textwidth]{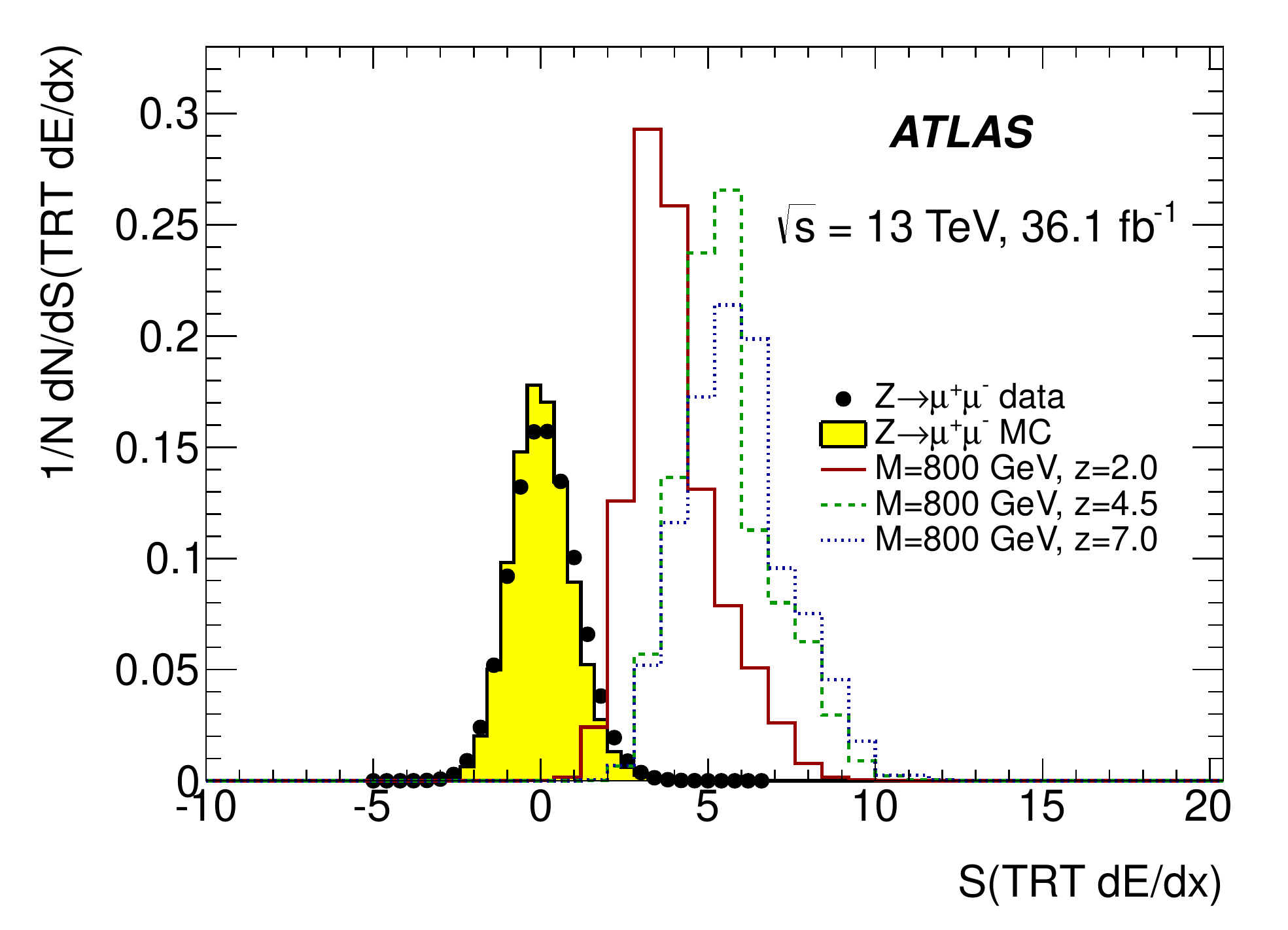}
   \includegraphics[width=0.49\textwidth]{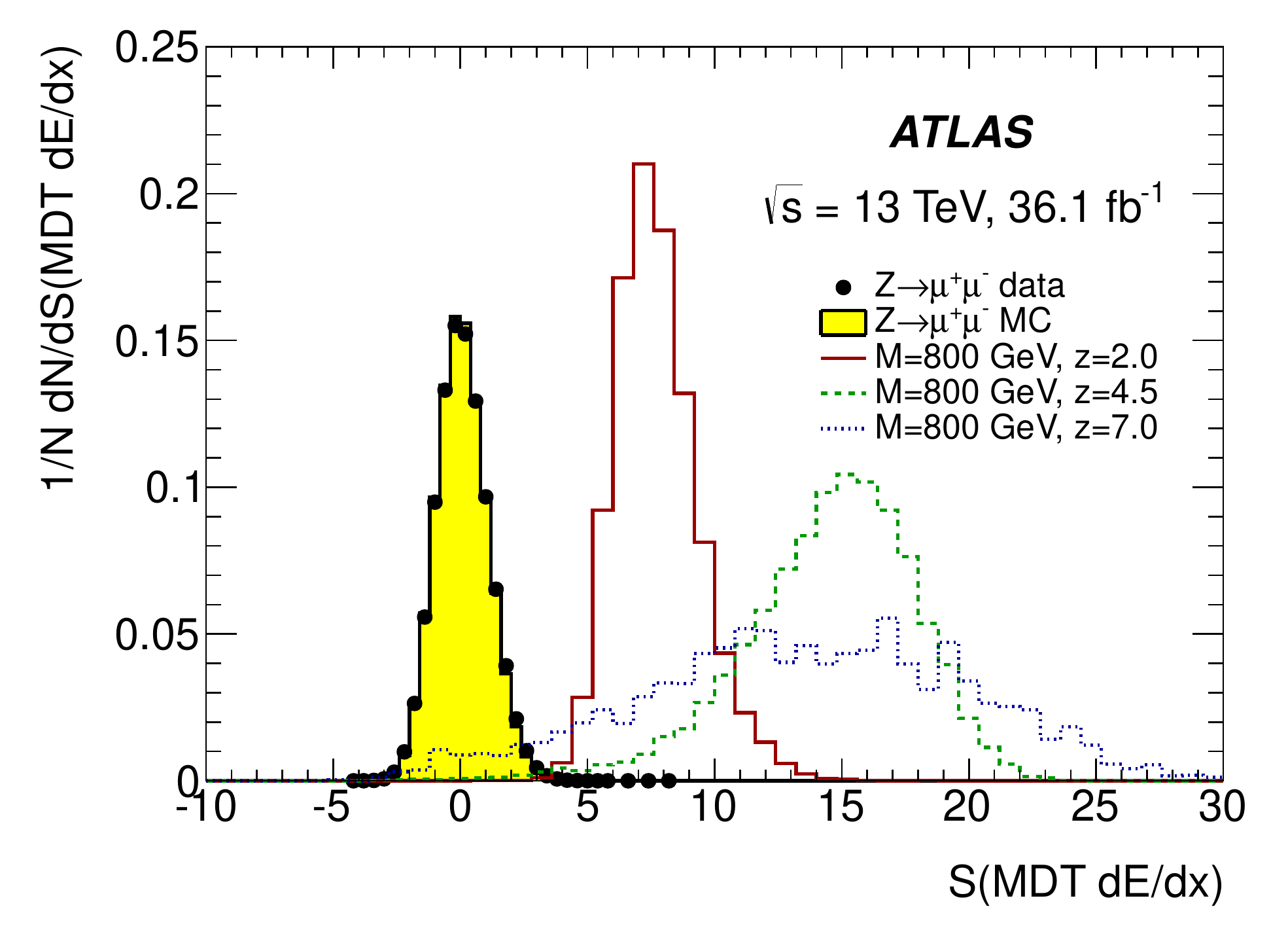}
 \end{center}
 \caption{Normalized distributions of the \dEdx{} significance in the TRT, $S$(TRT \dEdx{}), (left) 
   and in the MDT, $S$(MDT \dEdx{}), (right)
   for muons from $\Zmm{}$ events (data and simulation) and for simulated MCPs passing preselection requirements.
   Signal distributions are shown for ${z=2.0}$, $4.5$, and $7.0$, for a mass of \unit[800]{\GeV{}}.
   }
  \label{fig:dEdxSignificanceSeparation_MdtAndTrt}
\end{figure}

In the final selection, $S$(MDT \dEdx{}) and $S$(TRT \dEdx{})
are used as additional discriminating variables to separate signal from background.
\Fig{\ref{fig:dEdxSignificanceSeparation_MdtAndTrt}} shows the distributions of
these variables for muons from $\Zmm{}$ events compared with those expected from signal particles
with different charges (${z=2.0}$, $4.5$, and $7.0$) and a mass of \unit[800]{\GeV{}}.
It demonstrates that there is good separation between signal and background,
which increases with increasing charge.
The $S$(MDT \dEdx{}) distribution shape broadens noticeably with charge because, relative to typical muons, 
MCPs produce a larger number of $\delta$-rays, which give early-time hits in adjacent drift tubes. 
This results in the $\delta$-rays' ionization loss 
being measured instead of the MCP's loss, reducing the total ionization measured along the track.
The detailed detector response to these high-charge particles may not be well
simulated due to imperfect modeling of the saturation effects. However, since these two detectors (TRT and MDT)
do not lose signal at saturation, their most probable \dEdx{} values are higher than those of ${z=2}$
particles.

During the 2012 data-taking period, gas leaks started to develop in TRT pipes, located mostly in 
inaccessible areas, making their repair impossible~\cite{Aaboud:2017odu}. Due to the high cost of 
the xenon-based gas mixture, leaking modules were supplied with an argon-based mixture instead. 
The simulation does not fully model this change, resulting in a slightly narrower 
TRT \dEdx{} distribution in simulation than in data. This is a small effect because the ratio 
of signal amplitude to the low threshold for the argon-filled straws was tuned to be the same as for xenon-filled ones. 
It is accounted for in the systematic uncertainties calculation in \Sect{\ref{sec:EffUnc}}.

The energy loss in the calorimeters is not used in the search because they have coarser granularity than 
the tracking detectors and, thus, worse energy-loss resolution.  

Ionization-estimators discrepancies between data and simulation are accounted for as 
systematic uncertainties as described in \Sect{\ref{sec:EffUnc}}.

\begin{figure}[ht]
 \begin{center}
   \includegraphics[width=0.49\textwidth]{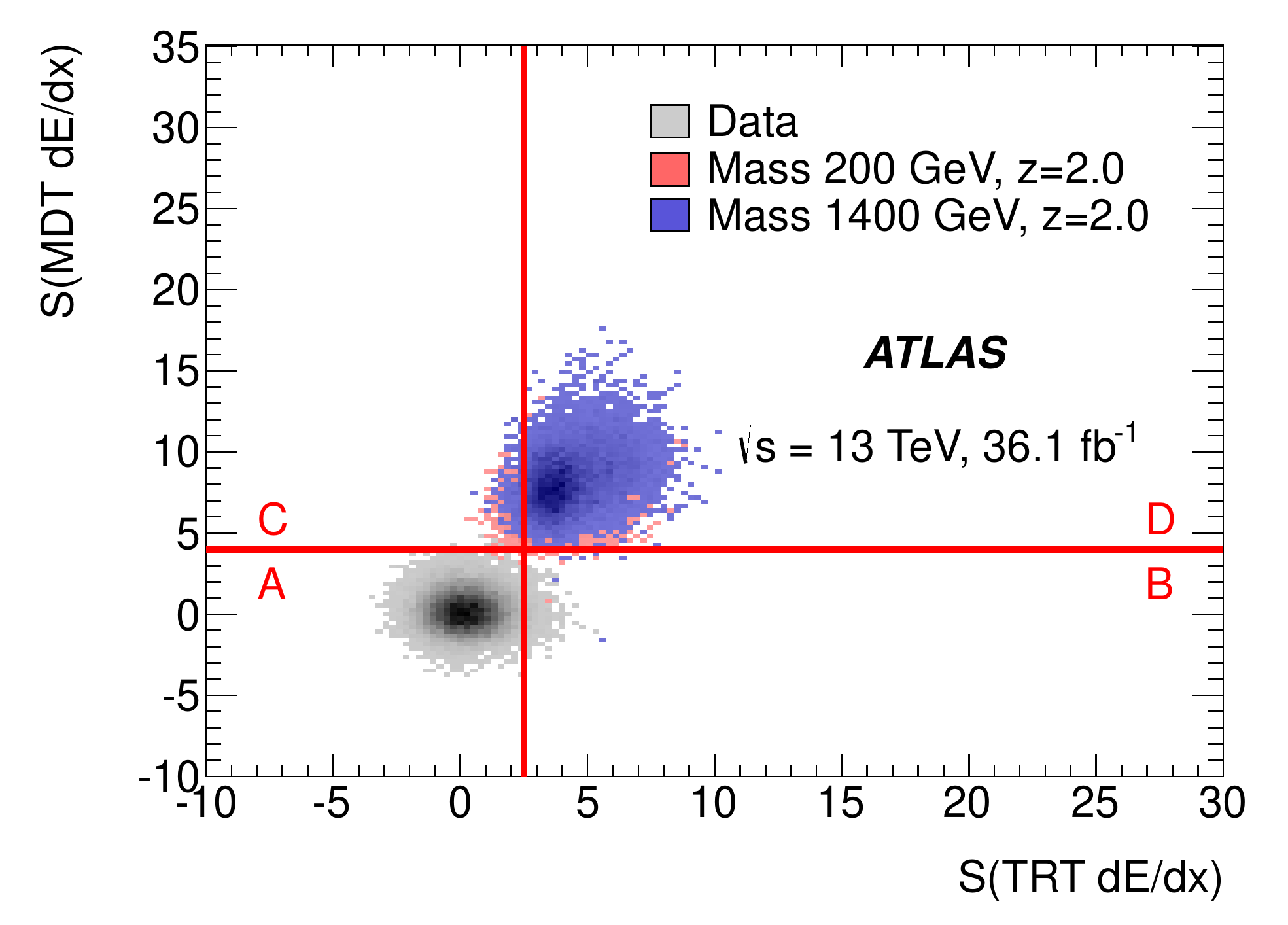}
   \includegraphics[width=0.49\textwidth]{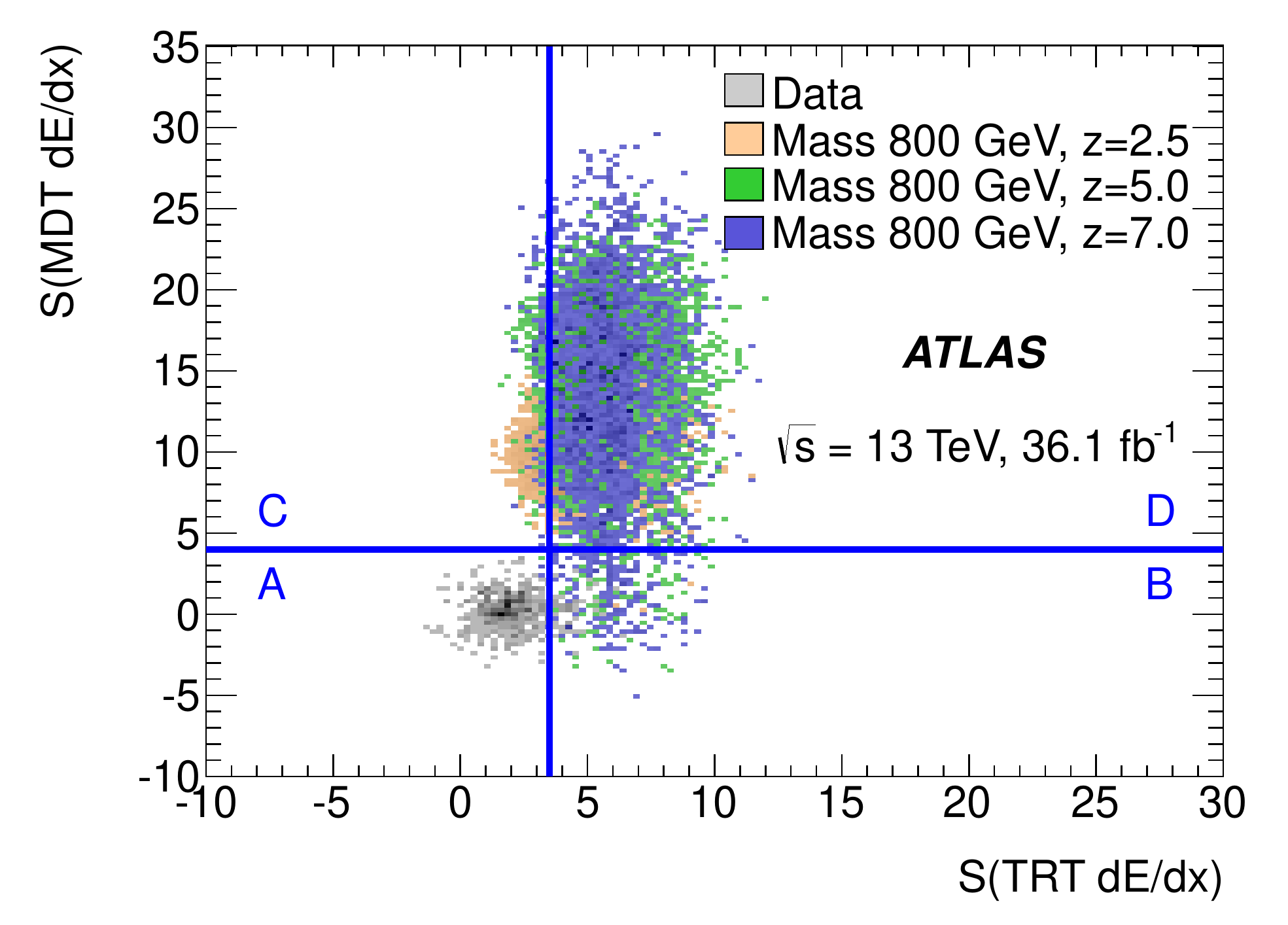}
 \end{center}
 \caption{$S$(MDT \dEdx{}) versus $S$(TRT \dEdx{}) after
   the ${z=2}$ (left) and ${z>2}$ (right) tight selections. The distributions of the data 
   and the simulated signal samples (for charges ${z=2.0}$, $2.5$, $5.0$, and $7.0$, and masses 
   of $200$, $800$, and \unit[1400]{\GeV{}}) are shown.
   The meaning of the A, B, C and D regions is discussed in the text.
  }
  \label{fig:AbcdPlanes}
\end{figure}

Two-dimensional distributions of $S$(MDT \dEdx{}) versus $S$(TRT \dEdx{})
are shown for data and simulated signal events in \Fig{\ref{fig:AbcdPlanes}} for candidates
passing the tight selection for ${z=2}$ (left) and ${z>2}$ (right).
As seen, the subdetector signatures are
different for the two samples, and thus the final signal
regions are chosen differently. They are defined by
$S$(TRT \dEdx{}) $> 2.5$ and $S$(MDT \dEdx{}) $> 4$ for
candidates selected as ${z=2}$ and by $S$(TRT \dEdx{}) $> 3.5$ and $S$(MDT \dEdx{}) $> 4$ 
for candidates selected as ${z>2}$.
The selection was optimized using only simulated
samples and $\Zmm{}$ data control samples without examining the signal region in the data.
 
A summary of the preselection, tight selection and final selection requirements for candidate tracks is 
presented in Table~\ref{tab:SelectionSummary}.

\begin{table}[htbp]
\caption{Summary of preselection, tight selection and final selection requirements.}
\begin{center}
\scalebox{0.9}{
\begin{tabular}{|c|c||c|c|c|}
\hline
\multicolumn{2}{|c||}{} & Candidate track & Tight & Final \\
\multicolumn{2}{|c||}{} & preselection & selection & selection \\
\hline
\multirow{10}{*}{\rot{Requirements}} & \multirow{5}{*}{\rot{${z=2}$}} & Combined muon with: & Preselected candidate with: & Tightly selected candidate with: \\
 & & & & \\
 & & \multicolumn{1}{c|}{``medium'' identification criteria,} & $S$(pixel \dEdx{}) $> 10$& $S$(TRT \dEdx{}) $> 2.5$,\\
 & & \multicolumn{1}{l|}{} & & $S$(MDT \dEdx{}) $> 4$\\
 & & \multicolumn{1}{c|}{$\pT^{\mu}/z > \unit[50]{\GeV{}},$} &  & \\
 \cline{2-2}\cline{4-5}
 & & & Preselected candidate with: & Tightly selected candidate with: \\
 & \multirow{3}{*}{\rot{${z>2}$}} & \multicolumn{1}{c|}{$\abseta < 2.0$,} &  & \\
 & & & \multicolumn{1}{ c| }{$\geq 1$ overflowing IBL cluster,} & \multicolumn{1}{ c| }{$S$(TRT \dEdx{}) $> 3.5$,} \\
 & & \multicolumn{1}{c|}{no other tracks with} & \multicolumn{1}{ c| }{$f^{\textrm{HT}} > 0.5$} & \multicolumn{1}{ c| }{$S$(MDT \dEdx{}) $> 4$} \\
 & & \multicolumn{1}{c|}{$\pT/z > \unit[0.5]{\GeV{}}$ within $\dR < 0.01$} &  &  \\
\hline
\end{tabular}
}
\end{center}
\label{tab:SelectionSummary}
\end{table}

\section{Expected background estimation}
\label{sec:BkgEstim}

The source of potential background is of an instrumental origin: it consists mainly of muons with ionization randomly 
fluctuating toward larger values due to detector occupancy effects (a large number of particles losing their 
energy in the same detector elements) and $\delta$-ray yields.
The expected background rate is estimated with two methods.

For the ${z=2}$ search case, it is estimated using an ABCD 
method~\cite{Aad:2013pqd}. According to this method, the plane of $S$(TRT \dEdx{}) and $S$(MDT \dEdx{})
is divided into regions A, B, C, and D using the final selection cuts
as shown in \Fig{\ref{fig:AbcdPlanes}}.
Region D is defined as the signal region, with regions A, B, and C
as control regions. 

If two or more particles contribute to the same event on the ABCD plane, only one is 
retained and shown on the plane according to a ``D-C-B-A-$\pT^{\mu}$'' ranking to avoid double-counting events. 
It consists in choosing the first particle found when considering region D then C, B, and finally A 
(ranked from the most populated quadrant to the least populated one in signal simulation).
If there are two or more particles in the event and in 
the same region, the highest-$\pT^{\mu}$ one is chosen.

For the ${z=2}$ search, the expected number of background events in the D region, $\ndata^{\textrm{D expected}}$, is estimated
from the numbers of observed data events in regions A, B, and C
($\ndata^{\textrm{A, B, C observed}}$):
 
\begin{equation*}
	\ndata^{\textrm{D expected}} = \frac{\ndata^{\textrm{B observed}}\times \ndata^{\textrm{C observed}}}{\ndata^{\textrm{A observed}}}.
	\label{eq:regularAbcd}
\end{equation*}

The same method is not used for the ${z>2}$ search case, because $\ndata^{\textrm{C observed}}=0$ 
(see \Fig{\ref{fig:AbcdPlanes}} (right)), which would lead to a large statistical uncertainty 
in $\ndata^{\textrm{D expected}}$. Instead, a method which employs sidebands of the two discriminating variables 
is used~\cite{Aad:2015oga}. Here, $\ndata^{\textrm{D expected}}$ is estimated from 
the number of observed events in region B and the probability $f$ to find
a particle with $S$(MDT \dEdx{}) $> 4$ in a single event:

\begin{equation*}
	\ndata^{\textrm{D expected}} = \ndata^{\textrm{B observed}}\times f.
	\label{eq:customAbcd}
\end{equation*}

\begin{figure}[ht]
 \begin{center}
   \includegraphics[width=0.6\textwidth]{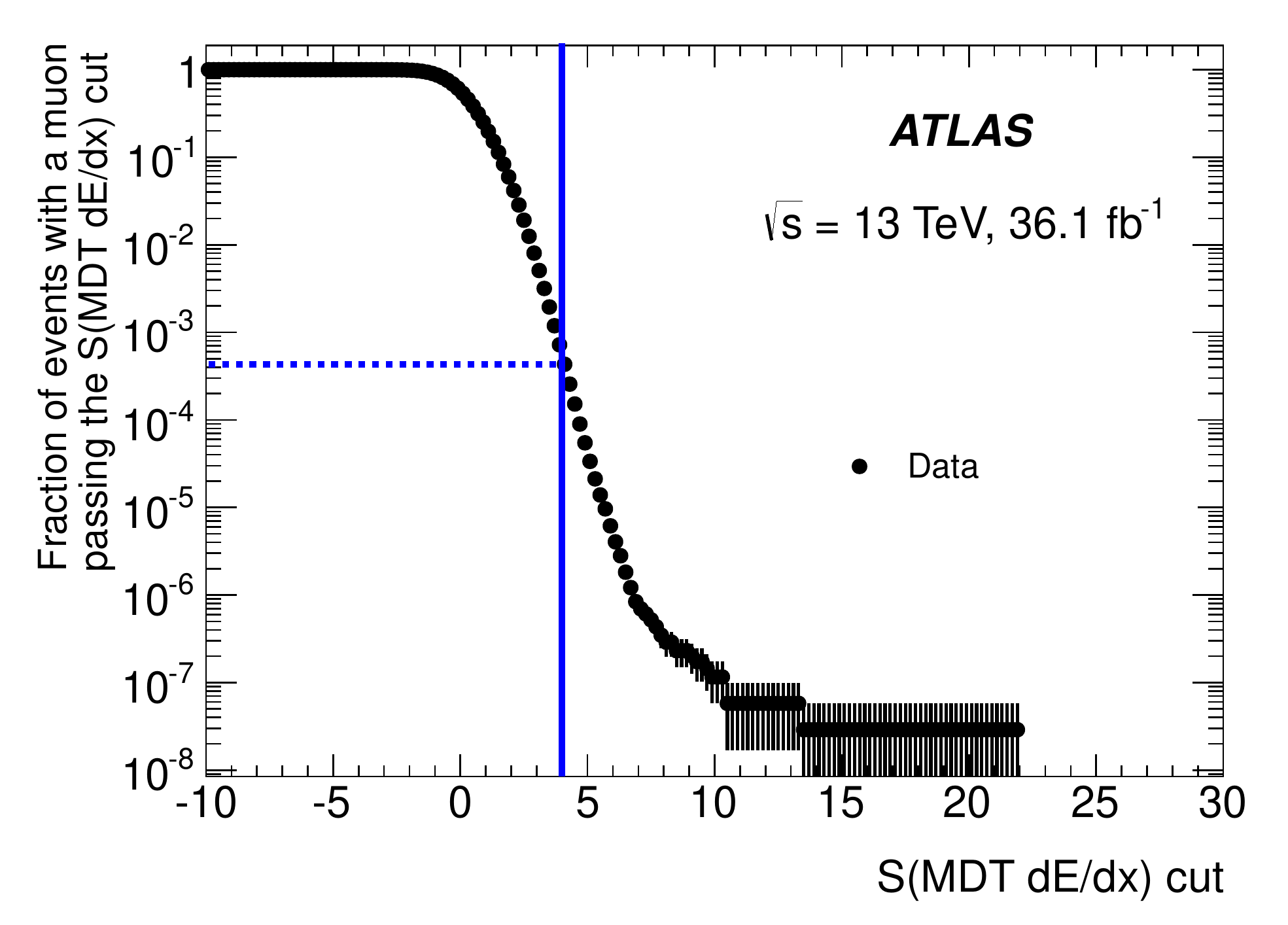}
 \end{center}
 \caption{Cumulative $S$(MDT \dEdx{}) distribution with the anti-tight selection applied, used to
   calculate the probability $f$ to find a particle with $S$(MDT \dEdx{})$ >4$. The resulting value of $f$ is 
   indicated by the blue dashed line.
  }
  \label{fig:MdtProb}
\end{figure}

The probability $f$ is derived from the cumulative $S$(MDT \dEdx{}) distribution for preselected muons in data with an anti-tight 
selection applied as shown in \Fig{\ref{fig:MdtProb}}. The anti-tight selection is defined by inverting one of the tight 
selection criteria: a muon must have $f^{\textrm{HT}} < 0.5$ or must not have any overflowing IBL clusters. 
If two or more muons are present in the same event, only the highest-$\pT^{\mu}$ muon is chosen to contribute to 
the distribution.

This method relies on the fact that $S$(MDT \dEdx{}) is not correlated with
the tight selection quantities in background events. 
A check was performed to demonstrate the absence of such correlations:
the distributions of $S$(pixel \dEdx{}), number of IBL clusters in overflow, $f^{\textrm{HT}}$ 
and $S$(TRT \dEdx{}) for muons in data with low $S$(MDT \dEdx{}) values (between $-10$ and $0$) 
were compared with those for muons with high $S$(MDT \dEdx{}) values (between $0$ and $10$). 
An agreement at a level of $98.5\%$ between the two cases is found, which shows that
there are no correlations between ionization estimators in different subdetectors
for background. Also, an additional check was performed to make sure that the shapes of cumulative 
$S$(MDT \dEdx{}) distributions for the cases of anti-tight (see \Fig{\ref{fig:MdtProb}}) and 
regular tight selections lay within their statistical uncertainties. Any residual differences are 
attributed to a systematic uncertainty as explained in \Sect{\ref{sec:BkgUnc}}.

The expected background contributions to the D regions and quantities used for their calculation
are shown in \Tabs{\ref{tab:yields_bkg_zEq2}}{\ref{tab:yields_bkg_zGt2}} for the ${z=2}$ and ${z>2}$ search 
cases, respectively. Systematic uncertainties in these values are estimated according to the method
discussed in \Sect{\ref{sec:BkgUnc}}.

\begin{table}[htbp]
	\caption{Expected background contribution (in events) to the D region in data for the ${z=2}$ selection, as well as quantities used for its calculation. The observed contribution is shown in the rightmost column.}
	\begin{center}
		\begin{tabular}{c c c c c}
			\hline
			$\ndata^{\textrm{A observed}}$  & $\ndata^{\textrm{B observed}}$  & $\ndata^{\textrm{C observed}}$  & $\ndata^{\textrm{D expected}}$   & $\ndata^{\textrm{D observed}}$ \\
			\hline
			$22117$ & $379$ & $9$ & $0.15 \pm 0.05$\stat{} $\pm \,0.10$\syst{} & $0$ \\
			\hline
		\end{tabular}
	\end{center}
	\label{tab:yields_bkg_zEq2}
\end{table}
 
\begin{table}[htbp]
	\caption{Expected background contribution (in events) to the D region in data for the ${z>2}$ selection, as well as quantities used for its calculation. The observed contribution is shown in the rightmost column.}
	\begin{center}
		\begin{tabular}{c c c c}
			\hline
			$\ndata^{\textrm{B observed}}$  & $f$ & $\ndata^{\textrm{D expected}}$ & $\ndata^{\textrm{D observed}}$ \\
			\hline
			$66$ & $4.3 \times 10^{-4}$  & ($2.9 \pm 0.4$\stat{} $\pm \,2.2$\syst{}) $\times 10^{-2}$ & $0$ \\
			\hline
		\end{tabular}
	\end{center}
	\label{tab:yields_bkg_zGt2}
\end{table}

In principle, the absence of candidates in the C region may be translated into a Poisson upper limit at $95\%$ 
confidence level of $2.996$~events, and the expected background yield can be further estimated with the ${z=2}$ method. 
However, this results in a less precise estimate with the upper limit of $0.5$~background events, 
which makes the usage of the ${z>2}$ method more favorable.

\section{Signal efficiency}
\label{sec:SigEff}

\begin{figure}[ht]
 \begin{center}
   \includegraphics[width=0.49\textwidth]{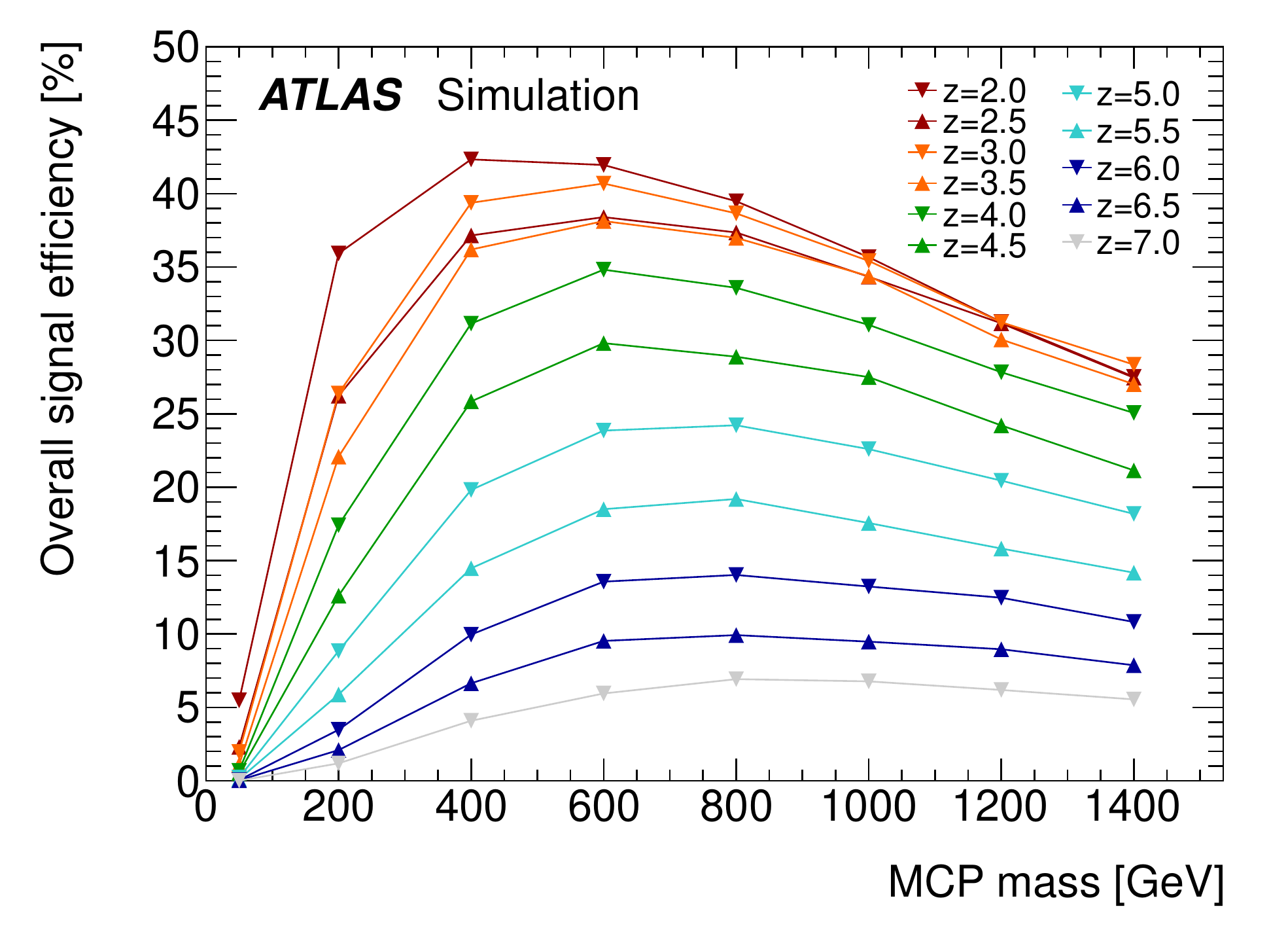}
   \includegraphics[width=0.49\textwidth]{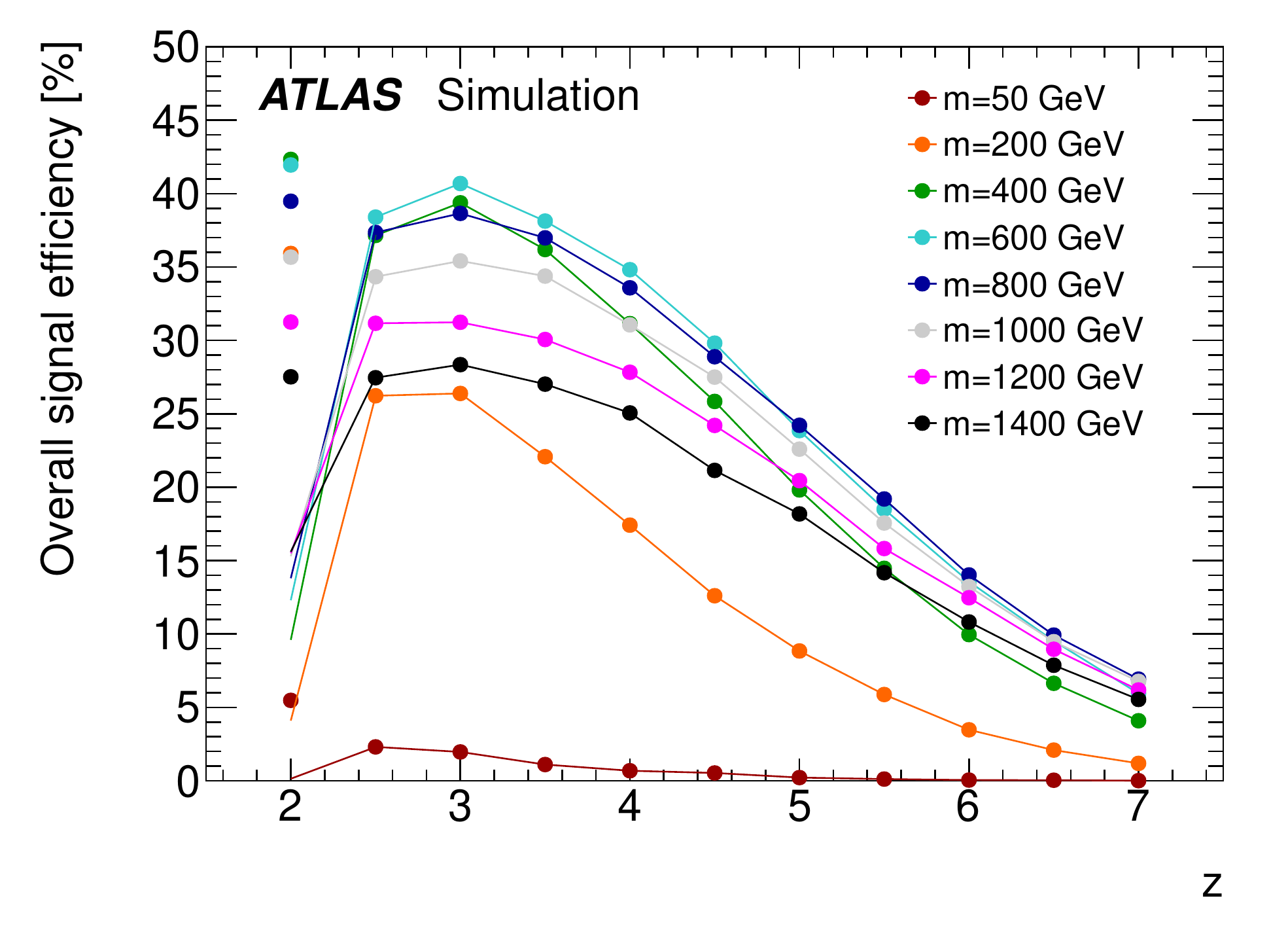}
 \end{center}
 \caption{
  The signal efficiencies for different MCP charges and masses for the DY production model versus, respectively, mass (left) and charge (right) values. Despite the analysis being performed separately for ${z=2}$ and ${z>2}$ MCPs, it is still sensitive to MCPs with ${2<z<2.5}$. In the right figure, the efficiencies at ${z=2.0}$ indicated by the continuous lines correspond to the efficiency values as if the ${z>2}$ selection was applied to the ${z=2}$ samples, thus denoting the conservative efficiency estimates for the ${2.0<z<2.5}$ particles. 
  }
  \label{fig:EfficiencyTrends}
\end{figure}

The cross-section limit is inversely proportional to the integrated luminosity of the analyzed data 
times the overall signal efficiency, which includes trigger and selection efficiencies.
This efficiency, as estimated from simulation, is shown in \Fig{\ref{fig:EfficiencyTrends}} 
for the signal samples used in this analysis.

The fraction of signal events satisfying the cumulative selection requirements is given 
in \Tab{\ref{tab:EfficiencyCutflow}} for several examples. 
 
\begin{table}[htbp]
	\caption{Fractions of signal events with at least one MCP candidate, which satisfy the given requirements (including all previous selection requirements). The uncertainties quoted are statistical only.}
	\begin{center}
		\begin{tabular}{|c|r|c|c|c|c|c|c|}
			\hline
			\multicolumn{2}{|c|}{Signal benchmark point} 	& Trigger 		& Candidate event 	& Tight 		& Final 		\\
			\cline{1-2}
			$z$	& Mass [$\GeV{}$]	 		& selection [\%]	& selection [\%]	& selection [\%]	& selection [\%]	\\
			\hline
			\hline
				& $200$ 	& $51.6\pm0.3$ & $41.6\pm0.3$ & $39.2\pm0.3$ & $35.9\pm0.3$ \\
			$2.0$ 	& $800$ 	& $54.5\pm0.3$ & $44.6\pm0.3$ & $41.3\pm0.3$ & $39.5\pm0.3$ \\
				& $1400$	& $39.9\pm0.3$ & $32.0\pm0.3$ & $28.4\pm0.3$ & $27.5\pm0.3$ \\
			\hline
				& $200$ 	& $19.2\pm0.2$ & $13.8\pm0.2$ & $13.4\pm0.2$ & $12.6\pm0.2$ \\
			$4.5$ 	& $800$		& $44.3\pm0.3$ & $31.3\pm0.3$ & $30.2\pm0.3$ & $28.9\pm0.3$ \\
				& $1400$	& $33.6\pm0.3$ & $22.6\pm0.2$ & $22.1\pm0.2$ & $21.1\pm0.2$ \\
			\hline
				& $200$		& $\ \:5.5\pm0.1$  	& $\ \:1.42\pm0.07$ & $\ \:1.4\pm0.1$ & $\ \:1.19\pm0.07$ \\
			$7.0$ 	& $800$ 	& $21.0\pm0.2$ 		& $\ \:8.3\pm0.2$   & $\ \:7.9\pm0.2$ & $\ \:6.9\pm0.2$   \\
				& $1400$	& $18.3\pm0.2$ 		& $\ \:6.7\pm0.1$   & $\ \:6.5\pm0.1$ & $\ \:5.5\pm0.1$   \\
			\hline
		\end{tabular}
	\end{center}
	\label{tab:EfficiencyCutflow}
\end{table}

Several factors contribute to the efficiency dependencies on mass and charge of the MCPs. For low masses, the ${\abseta<2.0}$ 
selection requirement and especially the $\pT^{\mu}/z$ requirement are the main sources of efficiency loss. 
This $\pT^{\mu}/z$ implied selection can be as high as approximately $\pT^{\mu}>50\times 7 = \unit[350]{\GeV{}}$, 
where $7$ is the highest charge value used in the analysis.
For high masses, the requirement to reach the MS with a velocity $\beta$ which satisfies the trigger timing window is 
the primary reason for the reduction in efficiency. 
Also, high ionization loss makes particles slow down: they may not fall within the trigger
timing window or may lose all their kinetic energy before reaching the MS.
The charge dependence of the efficiency results from higher ionization
and the higher effective $\pT^{\mu}/z$ selection, which are augmented by the factors $z^{2}$ and $z$, respectively.
Also, the increased production of $\delta$-rays at higher charges leads to a smaller number of reconstructed combined muons.
For events with only one MCP reaching the MS, the \met{} will be larger for heavier and/or 
higher-charged MCPs and therefore the \met{} trigger will be more likely to fire in such events.

\section{Uncertainties in the background estimation and signal yield}
\label{sec:Unc}

Uncertainties in the background estimate, the signal selection efficiency, and the integrated luminosity affect
the sensitivity of the search for MCPs. The contributions of these systematic uncertainties are described below.

\subsection{Background estimation uncertainty}
\label{sec:BkgUnc}

\begin{figure}[htbp]
	\begin{center}
		\includegraphics[width=0.49\textwidth]{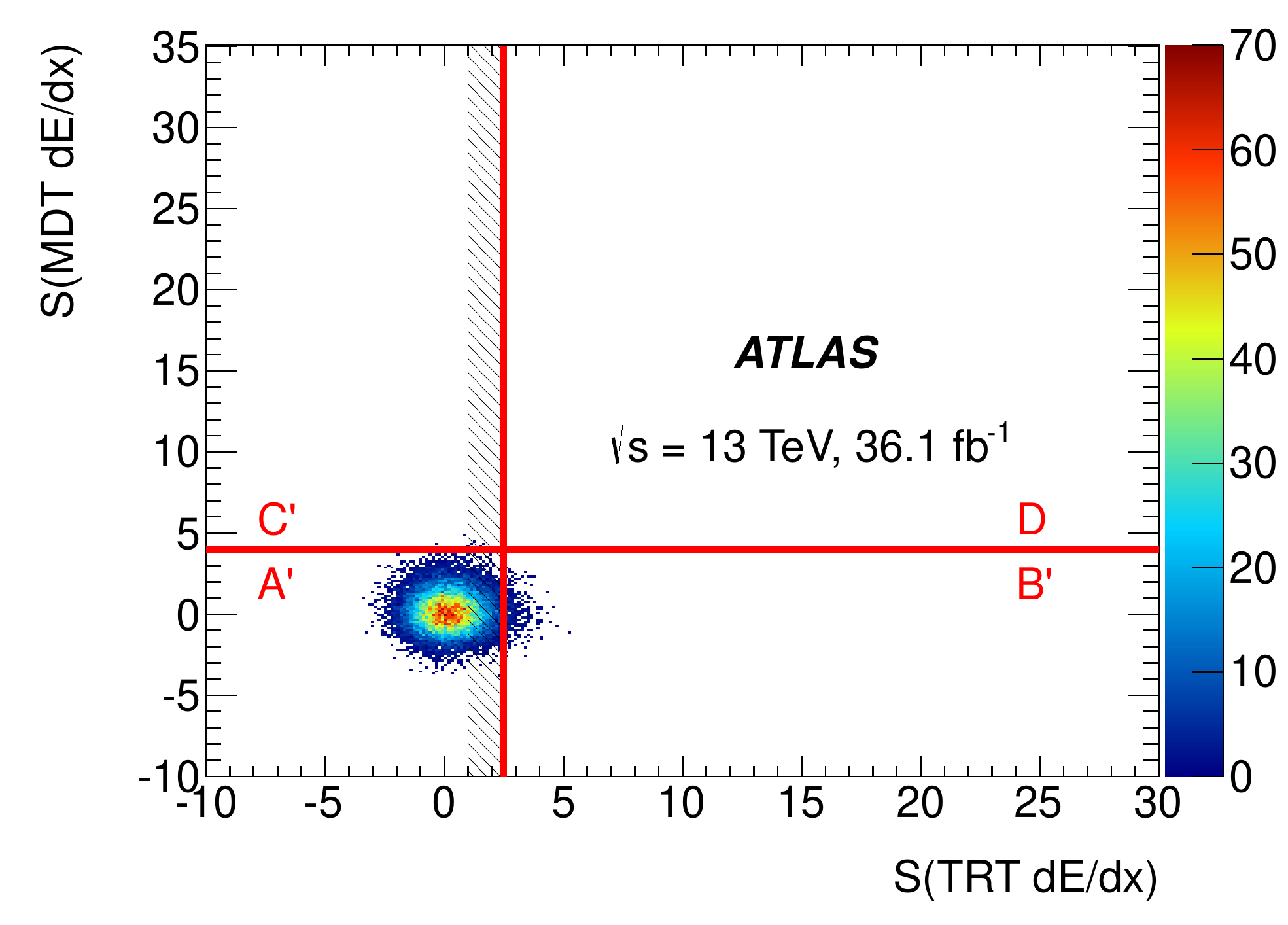}
		\includegraphics[width=0.49\textwidth]{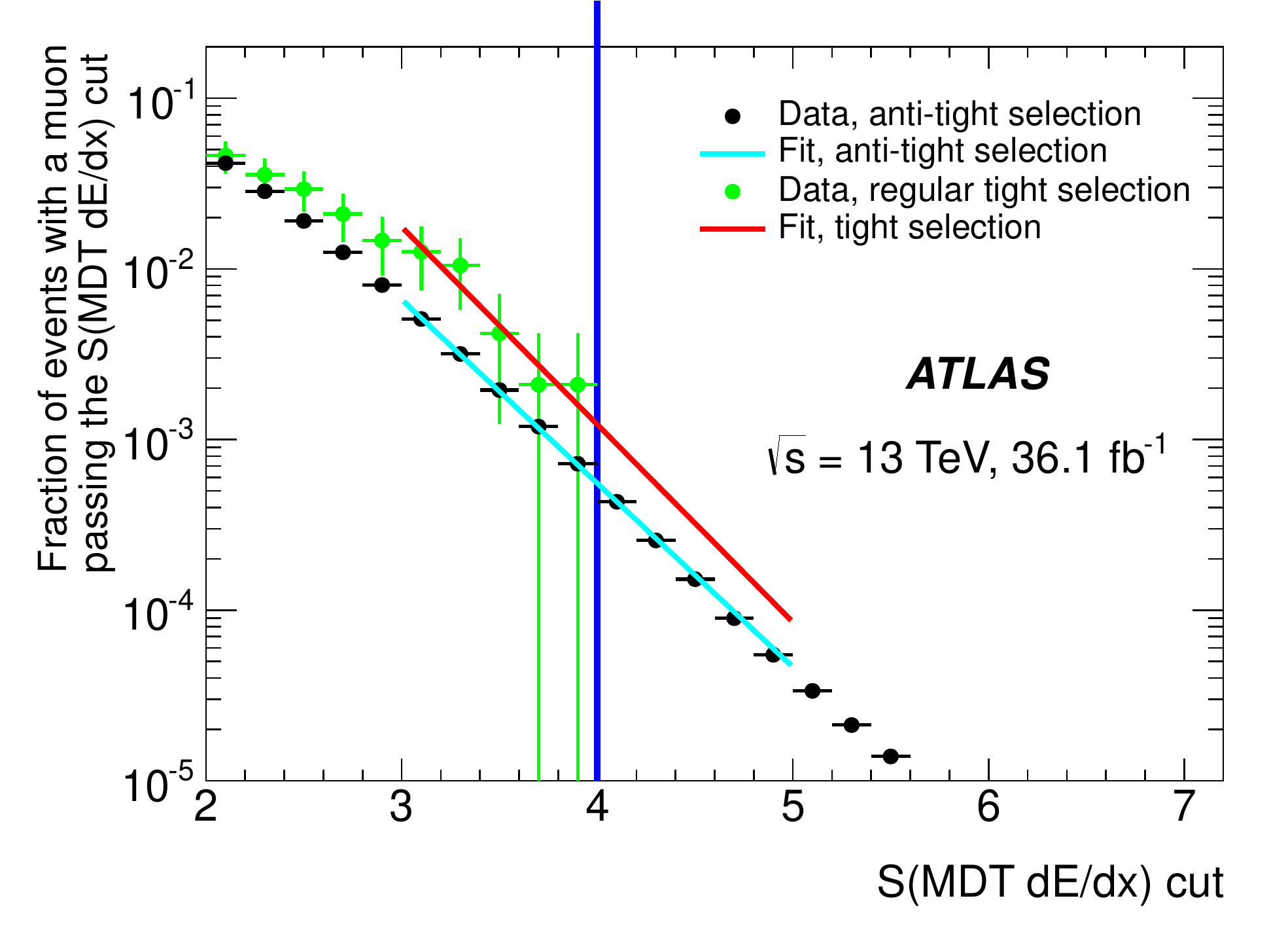}
	\end{center}
	\caption{Left: ABCD plane (for the ${z=2}$ case) used to assess the systematic uncertainty in the expected number of background events. Entries inside the ``dead region'' (here within $1.0< S$(TRT \dEdx{})$\:< 2.5$, shown by black shading) do not contribute to the background estimate used to assess the systematic uncertainty. Right: Overall $S$(MDT \dEdx{}) distribution with the anti-tight selection applied (black points) and with the regular tight selection applied (green points). Both distributions are shown zoomed in around the final selection cut (shown by a solid blue vertical line) and are fit with $p_0 \times {\textrm{e}}^{p_1x+p_2}$ functions to quantify their difference at $S$(MDT \dEdx{}) $=4$.
}
	\label{fig:BkgUncertainties}
\end{figure}

To assess a systematic uncertainty in the expected number of background events, so-called ``dead regions''
are introduced in the ABCD plane (see \Fig{\ref{fig:BkgUncertainties}}, left), and then the background estimate is recalculated
for several dead-region choices using the two methods described in \Sect{\ref{sec:BkgEstim}}. The dead regions used are:
$S_{\textrm{lower}}^{\textrm{MDT}} < S$(MDT \dEdx{}) $< 4.0$ with 
$S_{\textrm{lower}}^{\textrm{MDT}} = 2.0, 2.5, 3.0,$ and $3.5$ for both the ${z=2}$ and ${z>2}$ cases; and 
$S_{\textrm{lower}}^{\textrm{TRT}} < S$(TRT \dEdx{}) $< S_{\textrm{upper}}^{\textrm{TRT}}$ with 
$S_{\textrm{lower}}^{\textrm{TRT}} = 0.5, 1.0, 1.5,$ and $2.0$ for the ${z=2}$ case and 
$S_{\textrm{upper}}^{\textrm{TRT}} = 1.5, 2.0, 2.5,$ and $3.0$ for the ${z>2}$ case. 
The entries inside the dead regions do not contribute to the background estimate used to assess 
the systematic uncertainty. This method provides an insight into any possible correlations between 
the two variables used to construct the ABCD plane.

The maximum differences (calculated over several dead regions) between a new background expectation in the D 
region and the nominal one are $67\%$ ($0.1$~events) and $3.4\%$ ($10^{-3}$~events) for the ${z=2}$ and ${z>2}$ cases, 
respectively, and are treated as systematic uncertainties in the estimation of the expected background.

An additional uncertainty is assigned for the ${z>2}$ case due to a mismatch between the spectra of 
the fraction of events with a muon passing the $S$(MDT \dEdx{}) cut for the tight 
and anti-tight selections (see \Fig{\ref{fig:BkgUncertainties}}, right) at $S$(MDT \dEdx{}) values close to $4$. 
Both distributions were fit with a $p_0 \times {\textrm{e}}^{p_1x+p_2}$ function 
(a first-degree polynomial in the exponent was chosen for simplicity) within the range of $3< S$(MDT \dEdx{}) cut $< 5$, 
where $x$ represents the $S$(MDT \dEdx{}) cut value and $p_0$, $p_1$, and $p_2$ are free fit parameters. 
The difference between the two fits at $S$(MDT \dEdx{}) $=4$ 
is used to assign an additional $75\%$ systematic uncertainty ($0.022$~events). 

Summarizing the above, the final systematic uncertainties in the estimation of the expected background are $67\%$ for 
${z=2}$ and $75\%$ for ${z>2}$.

\subsection{Signal yield uncertainty}
\label{sec:EffUnc}

Several sources of systematic uncertainty in the signal efficiency are considered. 
The most significant uncertainties are those due to imperfect agreement between data and simulation, the trigger 
efficiency, and the parameterization of the parton distribution function used in the signal generation.

The uncertainty due to the disagreement between data and simulation is evaluated by varying
the signal acceptance requirements used in the analysis. Several considerations motivate these variations.
The uncertainty in the amount of material in front of the MS,
which is about $1\%$~\cite{Aad:2014rra}, 
propagates into an uncertainty in the selection efficiency due to the slowing down of particles,
and is covered by varying the $\pT^{\mu}$ requirement.
When considering the $\Zmm{}$ $\dEdx{}$ distributions together with those of the signal, the lower parts of 
the $\dEdx{}$ ranges are the most important for determining the signal efficiency. 
These correspond to the cores of the $\Zmm{}$ distributions and 
are the most relevant because if there is good agreement between data and simulation in that $\dEdx{}$ range 
(below the corresponding selection cut), the signal efficiencies will agree between the data and simulation. 
The variation applied to the nominal $\pT^{\mu}$ requirement is:
\begin{itemize}
\item $\pT^{\mu}$ value by $\pm3$\%. 
\end{itemize}

In addition, the mean and the root-mean-square width of the distributions in $\Zmm{}$ events disagree 
between data and simulation, and the ionization estimators may be mismodeled. These are accounted for by the following
variations of the signal selection criteria:

\begin{itemize}
\item number of overflowing IBL clusters by $\pm0.5\%$, 
\item $S$(pixel \dEdx{}) by $\pm25$\%, 
\item $f^{\textrm{HT}}$ by $\pm40$\%, 
\item $S$(TRT \dEdx{}) by $\pm15$\%, 
\item and $S$(MDT \dEdx{}) by $\pm3$\%.
\end{itemize}

The values of these variations are obtained by averaging the bin-by-bin ratios of $\Zmm{}$ yields in data to those in 
simulation (see \Figrange{\ref{fig:SPixeldEdx}}{\ref{fig:dEdxSignificanceSeparation_MdtAndTrt}}) in the cores 
of the corresponding distributions (within $\pm3\sigma$ with respect to the position of the mean of each distribution).
The total systematic uncertainties in the efficiency
arising from these variations range between $5\%$ and $80\%$, where the largest
uncertainty corresponds to lower-mass ${z=2.5}$ signal samples, which are fairly sensitive to 
the $f^{\textrm{HT}}$ variation.

The uncertainty in the trigger efficiency also has several sources, including 
an uncertainty in the muon-trigger efficiency (${<0.5\%}$), accounting for differences between 
triggering on the same muons in data and simulation,  
and an uncertainty in the \etmiss{} trigger efficiency ($23\%$ on average).
This second uncertainty depends on the accuracy of modeling the \etmiss{} turn-on curve, 
and is sensitive to the offline \etmiss{} reconstruction.
The former ($9.4\%$) was assessed by comparing the turn-on curves 
of the corresponding triggers in data and simulation using $\Zmm{}$ samples and taking the largest difference between all pairs.
The latter ($21\%$ on average) was assessed using the offline \etmiss{} spectra 
(in events triggered exclusively by the \etmiss{} trigger), varied to account for any possible uncertainties 
in the \etmiss{} term. 
There is also a $\beta$-dependent uncertainty originating from uncertainties in the modeling of the muon-trigger 
timing for particles with $\beta \ll 1$.
In order to improve the description of the trigger simulation, parameterized corrections 
were applied to the probability for MCPs to fire the RPC trigger.
To assess the uncertainty, the parameters of these corrections
were varied. The $\beta$ value of particles was varied between the true generated value
and the one reconstructed in the MS from the hypothesized mass and measured momentum. 
The time interval needed for a signal particle to reach the RPC trigger planes was varied by 
the root-mean-square width of the timing distribution for muons measured in the full $\Zmm{}$ data sample.
The uncertainty, assessed as the maximum relative difference between the nominal efficiency values and 
those obtained after the variations, ranges up to $1\%$ for signal particles with the highest charges and masses.
For the TGC trigger, no mismatch between the timing distributions in data and simulation was observed; 
therefore the trigger efficiency as obtained from simulation can be trusted.

For MCPs with $\beta$ significantly less than $1$ the drift time in the TRT and MDT could be mismeasured 
(due to the arrival time of the particle to the detector), worsening the momentum resolution, 
but TRT and MDT simulations model this effect. In the TRT, the effect is hardly noticeable due to the relatively 
small distance between the interaction point and the TRT. 
An MCP traveling one meter gets delayed by \unit[0.6]{ns} for $\beta=0.8$, 
while the TRT time bin is \unit[3.1]{ns}. 
Also, the track reconstruction accepts hits within a timing uncertainty of \unit[$\pm(3$--$4)$]{ns}. 
The time difference is larger for the MS (up to \unit[7]{ns}); however, the total drift time in MDT, with 
drift tubes of radius \unit[15]{mm}, is about \unit[700]{ns}, and thus any likely difference between 
data and simulation would not contribute significantly to the track reconstruction efficiency.

The NNPDF23LO parton distribution function (PDF) was varied within its error sets, each with a slightly different parameterization. 
These variations were translated into an uncertainty in the signal efficiency, ranging from $6\%$ for low-mass MCPs 
to $18\%$ for MCPs with the highest mass.

A $2.1\%$ uncertainty was assigned to the integrated luminosity used for this analysis. 
This uncertainty is derived, following a methodology 
similar to that detailed in \Ref{\cite{Aaboud:2016hhf}}, from a calibration of the luminosity scale 
using $x$--$y$ beam-separation scans. 

A subset of contributions from the separate sources of the most significant systematic uncertainties in 
the signal selection efficiency, as well as the resulting values of overall systematic uncertainties, 
are shown in \Tab{\ref{tab:EffUnc}} for several benchmark points. 

\begin{table}[htbp]
	\caption{Overview of the most significant individual contributions (in \%) to the overall systematic uncertainties in the signal selection efficiency, as well as the resulting values of the relative uncertainties (rightmost column).}
	\begin{center}
		\begin{tabular}{|c|r|c c c|c|}
			\hline
			\multicolumn{2}{|c|}{Signal benchmark point} & Data--simulation & Trigger effi- & PDF parameter- & Selection efficiency      \\
			\cline{1-2}
			$z$     & Mass [$\GeV{}$]                    & comparison [\%]  & ciency [\%]   & ization [\%]    & overall uncertainty [\%]  \\
			\hline
			\hline
				& $200$ & 12  		& \ \  0.9  	& \quad\  6.5 	& 14 \\
			$2.0$ 	& $800$ & \quad\  7.2 	& \ \  3.6  	& 10  		& 13 \\
				& $1400$& \quad\  6.1 	& \ \  5.0  	& 17  		& 19 \\
			\hline
				& $200$ & \quad\  8.9 	& \ \  1.4  	& \quad\  6.9 	& 12 \\
			$4.5$ 	& $800$ & \quad\  5.7 	& \ \  2.4  	& 11  		& 12 \\
				& $1400$& \quad\  5.9 	& 10.2 		& 17  		& 21 \\
			\hline
				& $200$ & \quad\  9.3 	& \ \  3.1  	& \quad\  7.2 	& 14 \\
			$7.0$ 	& $800$ & \quad\  6.6 	& \ \  8.2  	& 11  		& 16 \\
				& $1400$& \quad\  6.7 	& \ \  5.6  	& 18  		& 21 \\
			\hline
		\end{tabular}
	\end{center}
	\label{tab:EffUnc}
\end{table}

\section{Results}
\label{sec:Results}

No candidate events with MCPs were found for either the ${z=2}$ search or the ${z>2}$ search.
The results are consistent with the expectation of $0.15 \pm 0.05$\stat{} $\pm \,0.10$\syst{} and
($2.9 \pm 0.4$\stat{} $\pm \,2.2$\syst{}) $\times 10^{-2}$ background events, 
respectively. Since the number of events expected from background
is very small and no signal events were found, the observed and expected limits are practically identical. 

The limits are computed with the RooStats framework~\cite{2010acat.confE..57M}, 
which uses the CL$_\textrm{s}$ method~\cite{Read:2002hq} to discriminate
between the background-only hypothesis and the signal-plus-background hypothesis,
and determines exclusion limits for various MCP scenarios. The signal selection
efficiency, luminosity, expected and observed numbers of events 
and their uncertainties (as well as signal leakages -- fractions of the signal distributions 
outside the D region of the ABCD plane), 
handled as nuisance parameters,
are taken as input for pseudo-experiments, resulting in an observed limit at $95\%$ confidence level (CL). 

\begin{figure}[htbp]
	\begin{center}
		\includegraphics[width=0.6\textwidth]{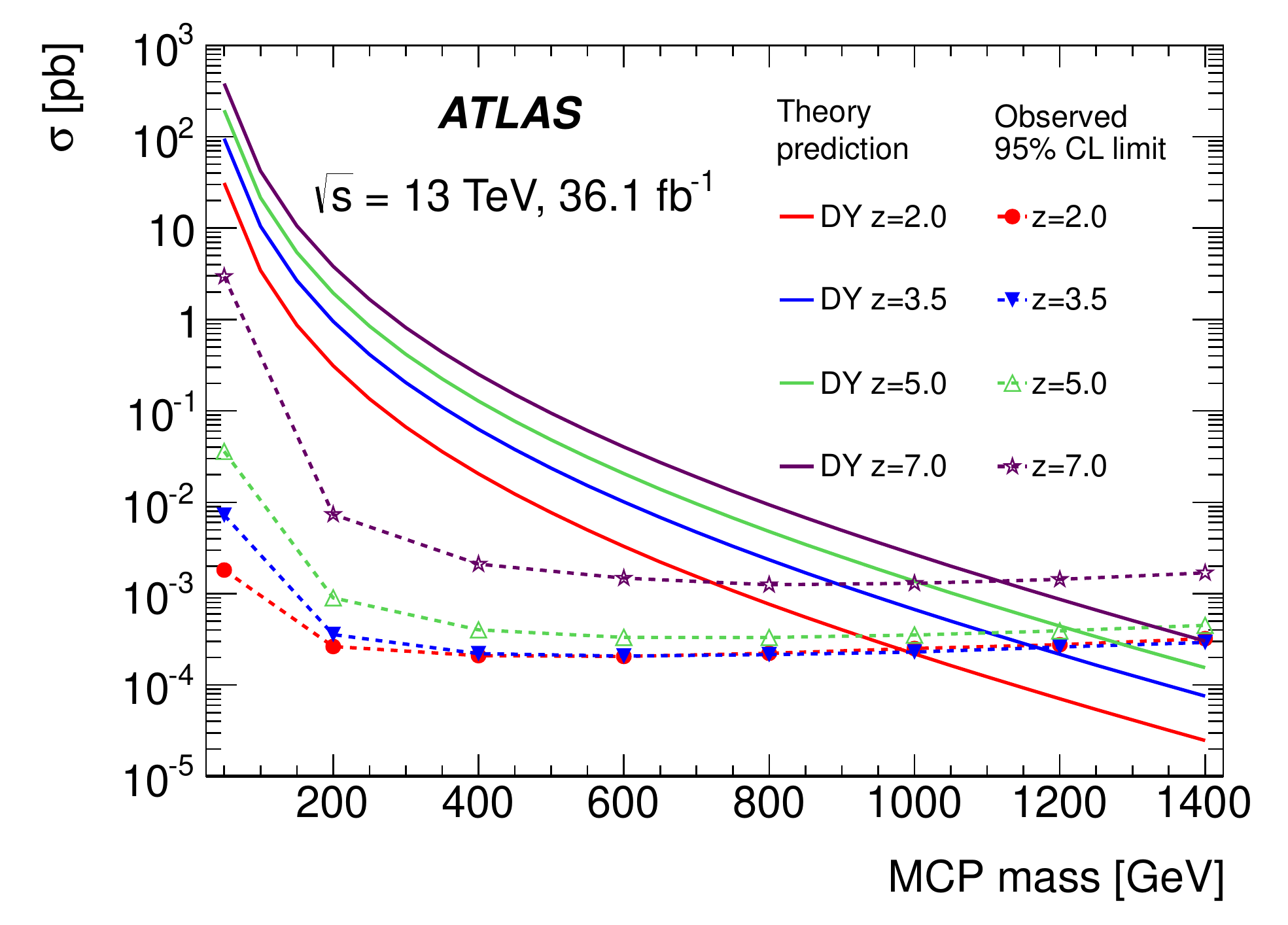}
	\end{center}
	\caption{Observed $95\%$ CL cross-section upper limits and theoretical cross-sections as functions of the lepton-like MCP's mass for several values of $z$ between $2$ and $7$.}
	\label{fig:limtrack}
\end{figure}

The measurement excludes the DY model of lepton-like MCP pair production over wide ranges of tested masses.
\Fig{\ref{fig:limtrack}} summarizes 
the observed $95\%$ CL cross-section limits as a function of mass for several MCP charges and compares them
with those predicted by the DY model.

\begin{figure}[htbp]
	\begin{center}
		\includegraphics[width=0.7\textwidth]{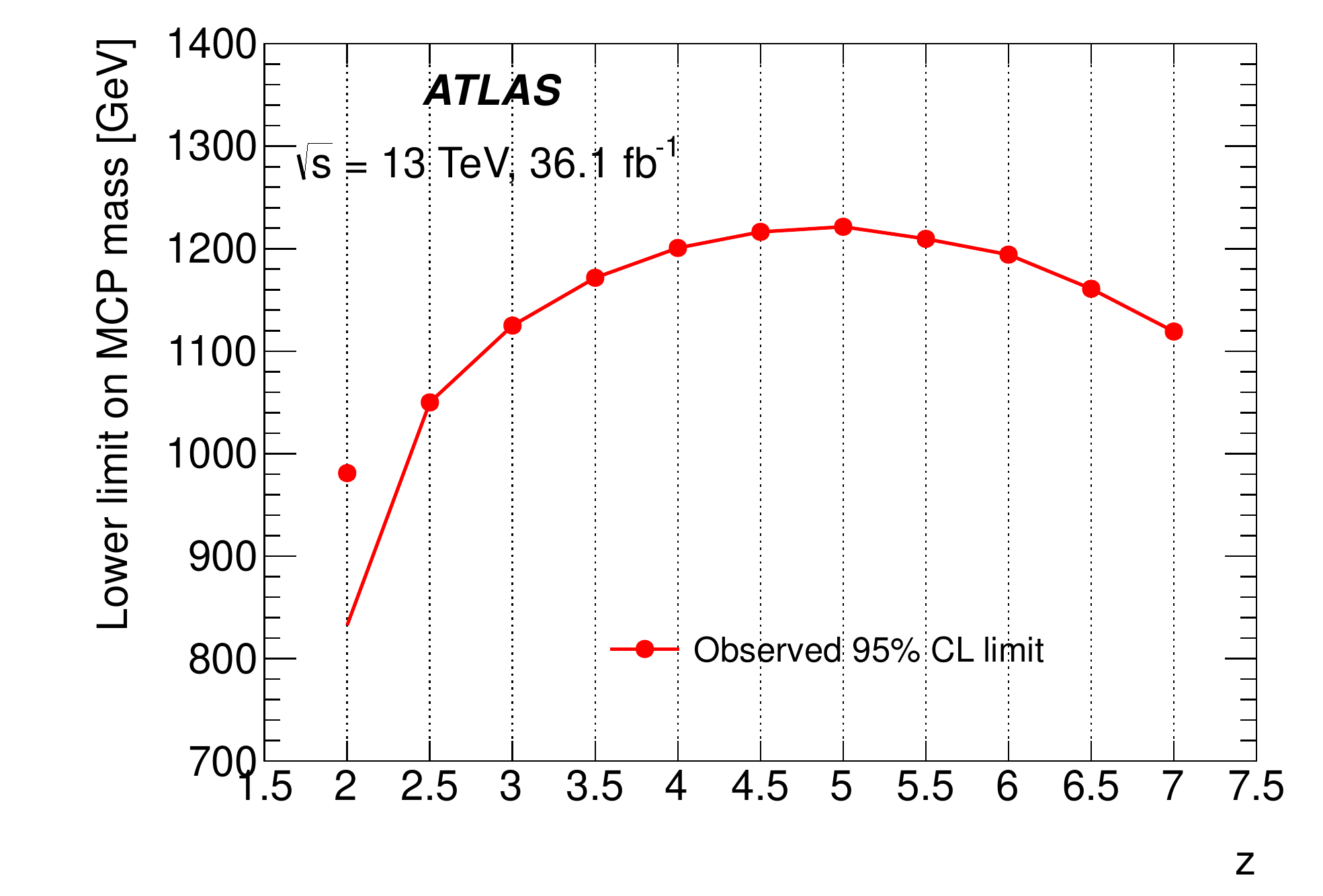}
	\end{center}
	\caption{Observed $95\%$ CL lower mass limits of lepton-like MCPs for charges $z \in [2,7]$ with a Drell--Yan pair-production model. The mismatch between the left end of the continuous line and the marker at $z=2.0$ (a shift by \unit[150]{\GeV{}}) is due to the difference in efficiencies between the cases where the ${z=2}$ and ${z>2}$ selections are applied to the ${z=2.0}$ samples, making this line segment constitute a conservative mass limit for $2.0<z<2.5$ particles.}
	\label{fig:MassExclusion}
\end{figure}

For this model, the cross-section limits can be transformed into
mass exclusion regions from \unit[50]{\GeV{}} up to the values in \Tab{\ref{tab:lowerMassLimits}}.
\Fig{\ref{fig:MassExclusion}} demonstrates the dependence of the lower mass exclusion limits on MCP charge values.
The mass limits are obtained from the intersection of the observed limits and the theoretical cross-section values.

\begin{table}[htbp]
	\caption{Observed $95\%$ CL lower mass limits of lepton-like MCPs for the Drell--Yan production model.}
	\begin{center}
		\begin{tabular}{|c|c c c c c c c c c c c|}
			\hline
			  & \multicolumn{11}{c|}{$z$} \\
			\cline{2-12}
			  & 2.0 & 2.5 & 3.0 & 3.5 & 4.0 & 4.5 & 5.0 & 5.5 & 6.0 & 6.5 & 7.0 \\
			\hline
			Lower mass limit [\TeV{}] & 0.98 & 1.06 & 1.13 & 1.17 & 1.20 & 1.22 & 1.22 & 1.21 & 1.19 & 1.16 & 1.12 \\
			\hline
		\end{tabular}
	\end{center}
	\label{tab:lowerMassLimits}
\end{table}

\FloatBarrier

\section{Conclusion}
\label{sec:Conclusion}

This article reports on a search for long-lived multi-charged particles
produced in proton--proton collisions with the ATLAS detector at the LHC.
The search uses a data sample with a center-of-mass energy of $\rts=\unit[13]{\TeV{}}$
and an integrated luminosity of \intlumi{}. Lepton-like particles are searched for with electric
charges from ${|q|=2e}$ to ${|q|=7e}$ penetrating the full ATLAS detector
and producing anomalously high ionization signals in multiple detector
elements. Less than one background event is expected and
no events are observed. Upper limits are derived on the cross-sections
using a Drell--Yan production model and 
exclude lepton-like multi-charged particles with masses between $\unit[50]{\GeV{}}$ and $980$--$\unit[1220]{\GeV{}}$.

\FloatBarrier

\section*{Acknowledgments}

We thank CERN for the very successful operation of the LHC, as well as the
support staff from our institutions without whom ATLAS could not be
operated efficiently.

We acknowledge the support of ANPCyT, Argentina; YerPhI, Armenia; ARC, Australia; BMWFW and FWF, Austria; ANAS, Azerbaijan; SSTC, Belarus; CNPq and FAPESP, Brazil; NSERC, NRC and CFI, Canada; CERN; CONICYT, Chile; CAS, MOST and NSFC, China; COLCIENCIAS, Colombia; MSMT CR, MPO CR and VSC CR, Czech Republic; DNRF and DNSRC, Denmark; IN2P3-CNRS, CEA-DRF/IRFU, France; SRNSFG, Georgia; BMBF, HGF, and MPG, Germany; GSRT, Greece; RGC, Hong Kong SAR, China; ISF and Benoziyo Center, Israel; INFN, Italy; MEXT and JSPS, Japan; CNRST, Morocco; NWO, Netherlands; RCN, Norway; MNiSW and NCN, Poland; FCT, Portugal; MNE/IFA, Romania; MES of Russia and NRC KI, Russian Federation; JINR; MESTD, Serbia; MSSR, Slovakia; ARRS and MIZ\v{S}, Slovenia; DST/NRF, South Africa; MINECO, Spain; SRC and Wallenberg Foundation, Sweden; SERI, SNSF and Cantons of Bern and Geneva, Switzerland; MOST, Taiwan; TAEK, Turkey; STFC, United Kingdom; DOE and NSF, United States of America. In addition, individual groups and members have received support from BCKDF, CANARIE, CRC and Compute Canada, Canada; COST, ERC, ERDF, Horizon 2020, and Marie Sk{\l}odowska-Curie Actions, European Union; Investissements d' Avenir Labex and Idex, ANR, France; DFG and AvH Foundation, Germany; Herakleitos, Thales and Aristeia programmes co-financed by EU-ESF and the Greek NSRF, Greece; BSF-NSF and GIF, Israel; CERCA Programme Generalitat de Catalunya, Spain; The Royal Society and Leverhulme Trust, United Kingdom. 

The crucial computing support from all WLCG partners is acknowledged gratefully, in particular from CERN, the ATLAS Tier-1 facilities at TRIUMF (Canada), NDGF (Denmark, Norway, Sweden), CC-IN2P3 (France), KIT/GridKA (Germany), INFN-CNAF (Italy), NL-T1 (Netherlands), PIC (Spain), ASGC (Taiwan), RAL (UK) and BNL (USA), the Tier-2 facilities worldwide and large non-WLCG resource providers. Major contributors of computing resources are listed in Ref.~\cite{ATL-GEN-PUB-2016-002}.

\clearpage
\printbibliography

\clearpage
 
\begin{flushleft}
{\Large The ATLAS Collaboration}

\bigskip

M.~Aaboud$^\textrm{\scriptsize 34d}$,    
G.~Aad$^\textrm{\scriptsize 99}$,    
B.~Abbott$^\textrm{\scriptsize 125}$,    
D.C.~Abbott$^\textrm{\scriptsize 100}$,    
O.~Abdinov$^\textrm{\scriptsize 13,*}$,    
B.~Abeloos$^\textrm{\scriptsize 129}$,    
D.K.~Abhayasinghe$^\textrm{\scriptsize 91}$,    
S.H.~Abidi$^\textrm{\scriptsize 164}$,    
O.S.~AbouZeid$^\textrm{\scriptsize 39}$,    
N.L.~Abraham$^\textrm{\scriptsize 153}$
% ...
\bigskip
\\

% ...
$^{13}$Institute of Physics, Azerbaijan Academy of Sciences, Baku; Azerbaijan.\\
% ...
$^{34}$$^{(a)}$Facult\'e des Sciences Ain Chock, R\'eseau Universitaire de Physique des Hautes Energies - Universit\'e Hassan II, Casablanca;$^{(b)}$Centre National de l'Energie des Sciences Techniques Nucleaires (CNESTEN), Rabat;$^{(c)}$Facult\'e des Sciences Semlalia, Universit\'e Cadi Ayyad, LPHEA-Marrakech;$^{(d)}$Facult\'e des Sciences, Universit\'e Mohamed Premier and LPTPM, Oujda;$^{(e)}$Facult\'e des sciences, Universit\'e Mohammed V, Rabat; Morocco.\\
% ...
$^{39}$Niels Bohr Institute, University of Copenhagen, Copenhagen; Denmark.\\
% ...
$^{91}$Department of Physics, Royal Holloway University of London, Egham; United Kingdom.\\
% ...
$^{99}$CPPM, Aix-Marseille Universit\'e, CNRS/IN2P3, Marseille; France.\\
$^{100}$Department of Physics, University of Massachusetts, Amherst MA; United States of America.\\
% ...
$^{125}$Homer L. Dodge Department of Physics and Astronomy, University of Oklahoma, Norman OK; United States of America.\\
% ...
$^{129}$LAL, Universit\'e Paris-Sud, CNRS/IN2P3, Universit\'e Paris-Saclay, Orsay; France.\\
% ...
$^{153}$Department of Physics and Astronomy, University of Sussex, Brighton; United Kingdom.\\
% ...
$^{164}$Department of Physics, University of Toronto, Toronto ON; Canada.\\
% ...
$^{a}$ Also at Borough of Manhattan Community College, City University of New York, NY; United States of America.\\
$^{b}$ Also at California State University, East Bay; United States of America.\\
$^{c}$ Also at Centre for High Performance Computing, CSIR Campus, Rosebank, Cape Town; South Africa.\\
$^{d}$ Also at CERN, Geneva; Switzerland.\\
$^{e}$ Also at CPPM, Aix-Marseille Universit\'e, CNRS/IN2P3, Marseille; France.\\
% ...
$^{*}$ Deceased

\end{flushleft}


\end{document}